\documentclass[aps,prb,floatfix,twocolumn,preprintnumbers,amsmath,amssymb,groupedaddress,showpacs,showkeys]{revtex4-1}

\usepackage{graphicx}  
\usepackage{dcolumn}   
\usepackage{bm}        
\usepackage{color}     
\usepackage{amsmath,amssymb}
\usepackage{psfrag}
\usepackage{nicefrac}
\usepackage{wasysym}

\usepackage{braket}  

\newcommand{\llangle}{\langle\hspace{-0.4mm}\langle}
\newcommand{\rrangle}{\rangle\hspace{-0.4mm}\rangle}
\newcommand{\ua}{\uparrow}
\newcommand{\da}{\downarrow}

\newcommand{\lI}{\lambda_{\rm{I}}}

\newcommand{\LIA}{\Lambda^{\rm{A}}_{\rm{I}}}
\newcommand{\LIB}{\Lambda^{\rm{B}}_{\rm{I}}}
\newcommand{\LBR}{\Lambda_{\rm{R}}}
\newcommand{\LFC}{\Lambda_{\rm{FC}}}

\newcommand{\LPA}{\Lambda^{\rm{A}}_{\rm{PIA}}}
\newcommand{\LPB}{\Lambda^{\rm{B}}_{\rm{PIA}}}
\newcommand{\s}{\sigma}
\newcommand{\ii}{\mathrm{i}}

\begin{document}

\title{Spin-orbit coupling in fluorinated graphene}
\author{Susanne Irmer, Tobias Frank, Sebastian Putz, Martin Gmitra, Denis Kochan, and Jaroslav Fabian}
\affiliation{Institute for Theoretical Physics, University of Regensburg, 93040 Regensburg, Germany\\
}
\begin{abstract}
We report on theoretical investigations of the spin-orbit coupling effects in fluorinated graphene. First-principles
density functional calculations are performed for the dense and dilute adatom coverage limits.
The dense limit is represented by the single-side semifluorinated graphene, which is a metal with spin-orbit
splittings of about 10 meV. To simulate the effects of a single adatom, we also calculate the electronic
structure of a $10 \times 10$ supercell, with one fluorine atom in the top position. Since this dilute
limit is useful to study spin transport and spin relaxation, we also introduce a realistic effective hopping Hamiltonian, based
on symmetry considerations, which describes the supercell bands around the Fermi level. We provide the
Hamiltonian parameters which are best fits to the first-principles data. We demonstrate that, unlike for
the case of hydrogen adatoms, fluorine's own spin-orbit coupling is the principal cause of the giant induced
local spin-orbit coupling in graphene. The $sp^3$ hybridization induced transfer of spin-orbit coupling from
graphene's $\sigma$ bonds, important for hydrogenated graphene, contributes
much less. Furthermore, the magnitude of the induced spin-orbit coupling due to fluorine adatoms is
about $1000$ times more than that of pristine graphene, and 10 times more than that of hydrogenated graphene.
Also unlike hydrogen, the fluorine adatom is not a narrow resonant scatterer at the Dirac point.
The resonant peak in the density of states of fluorinated graphene in the dilute limit
lies 260 meV below the Dirac point. The peak is rather broad, about 300 meV, making the fluorine adatom
only a weakly resonant scatterer.
\end{abstract}

\pacs{}
\keywords{}
\date{\today}
\maketitle

\section{Introduction}

Functionalizing graphene with adatoms and admolecules is very attractive for
spintronics research,\cite{Zutic2004:RMP} as it offers new tools to tailor spin and magnetic properties
\cite{Han2014:NatCom} of this unique two-dimensional material.\cite{Novoselov2012:Nat}
On the one hand, adatoms---in particular paramagnetic ones such as hydrogen \cite{Yazyev2010:RPP} and
organic molecules \cite{Santos2012:NJP}---can provide the mechanism \cite{Kochan2014:PRL,Tuan2014:NP}
for the ultrafast spin relaxation seen in spin injection experiments.\cite{Tombros2007:Nat, Yang2011:PRL, Han2011:PRL, Dlubak2012:NP}
On the other hand, even light adatoms such as hydrogen can induce giant spin-orbit
coupling (SOC)\cite{Neto2009:PRL, Gmitra2013:PRL}
which makes novel spin transport phenomena such as the spin Hall effect \cite{Balakrishnan2013:NP} observable in graphene.

A particularly interesting adatom is fluorine. Being the most electronegative element,
fluorine forms a strong covalent bond with carbon, affecting the stability of possible
fluorine-graphene conformations.\cite{Peeters2010:PRB} Even before the synthesis
of graphene there were theoretical studies of fluorinated graphite surfaces.\cite{Charlier1993:PRB, Takagi2002:PRB}
Magnetization measurements show that fluorinated graphene can induce spin 1/2 paramagnetic moments.\cite{Nair2012:NP}
Magnetic moments also seem to be deduced from magnetotransport \cite{Hong2011:PRB} and
weak localization measurements.\cite{Hong2012:PRL}
However, theoretically the case for induced magnetic moments due to fluorine is
controversial. Some density functional calculations predict a spin unpolarized
ground state,\cite{Santos2012:NJP} while other calculations seem to predict a spin polarized
one.\cite{Kim2013:PRB} The difficulty seems to stem from
the self-interaction error in the exchange-correlation functionals that tends to
delocalize electronic states.\cite{MoriSanchez2008:PRL, Casolo2010:PRB}
It may be that the standard density functional approximations will not be able
to resolve this case, calling perhaps for more advanced quantum chemistry approaches. There are also investigations of the influence of charge doping on the magnetism in fluorine. \cite{Yndurain2013:PRB} These studies show that charge doping not only affects the magnetism in fluorinated graphene but also leads to a transition from covalent to ionic bonding.\cite{Sofo2011:PRB}

Here we focus on spin-orbit coupling induced by fluorine adatoms in graphene.
We perform
first-principles density functional calculations in the dense and dilute limits of fluorine
coverage, to quantify the spin-orbit splitting of the relevant energy bands. Our dense limit is given by the single-side semifluorinated graphene ($\mathrm{C_2F}$) which can
be experimentally prepared by the chemical reduction of oxidized graphite surfaces.\cite{Okotrub2009:PSSB}
Our study focuses on the chair conformation of ${\rm C_2F}$ which is metallic within DFT, as well as within GW
calculations.\cite{Sahin2011:PRB} The computed phonon spectrum also shows the dynamical stability of this
structure.\cite{Sahin2011:PRB} It is predicted that the
ground state of ${\rm C_2F}$ should be a N\'{e}el antiferromagnet.\cite{Rudenko2013:PRB}

Our main finding is a giant enhancement of spin-orbit coupling, compared even to hydrogenated
graphene.\cite{Gmitra2013:PRL} The spin-orbit splitting reaches 30 meV. Such a splitting cannot come from graphene
itself, as the $\sigma$ bonds are split by about 10 meV, which is the native spin-orbit splitting of the
carbon atom. Instead, the splitting comes from the fluorine adatom. This conclusion is further
confirmed by investigating the dilute limit, represented here by a $10\times 10$ supercell.
For this limit we also derive an effective minimal hopping orbital and spin-orbit Hamiltonian, with
parameters obtained by fitting to our first-principles calculations. This realistic Hamiltonian
should be useful for model calculations of spin transport and spin relaxation in
fluorinated graphene. In particular, it shows that fluorine adatoms in the top position
are not resonant scatterers at the Dirac point. We present tight-binding calculations
for a large, $40 \times 40$, supercell, as well as analytical results for a single adatom,
to show that fluorine leads to a very broad, indeed weak or marginal resonant scattering
off the Dirac point: the peak
lies 260 meV below the Dirac point, having the full width at half maximum of 300 meV.

Our first-principles results, for both the dense and dilute limits, were performed with
the full potential linearized augmented plane wave method, within density functional theory (DFT).
We work with spin unpolarized ground states, to identify spin-orbit effects in the band structures.
To model different concentrations of F adatoms we chose different supercells.
The dense ($1 \times 1$ supercell) and intermediate ($5 \times 5$ supercell) coverages were calculated with the WIEN2K code,\cite{wien2k} using a vacuum spacing of 15~\AA.
The dilute limit ($10 \times 10$ supercell) was calculated with FLEUR \cite{Fleur} in the film
geometry. Spin-orbit coupling in these codes is included within the second-variational step in
the muffin-tin spheres for valence electrons, while the core electrons are treated fully relativistically.
All our first-principles calculations are well converged for fully relaxed structures.

The paper is organized as follows. In Sec.~II we present DFT results of the electronic band structure calculation and
specifically the spin-orbit coupling induced spin splitting of the bands for a semifluorinated graphene (50\% coverage), representing
our dense limit case. Section~III\,A brings the same, but for a larger supercell, $10 \times 10$ supercell (0.5\% coverage) representing our dilute limit.
The dilute limit is then analyzed using a symmetry derived effective hopping Hamiltonian in Sec.~III\,B.
The model parameters of the Hamiltonian are fitted to the first-principles results. In Sec.~III\,C we
answer the question whether or not fluorine adatoms on graphene are resonant scatterers.
Finally, a comparative analysis with hydrogenated graphene is provided in the Appendix,
using an intermediate size $5\times 5$ supercell.

\section{Dense limit: single-side semifluorinated graphene $\mathrm{C_2F}$}

\begin{figure}[htp]
 \includegraphics[width = \columnwidth]{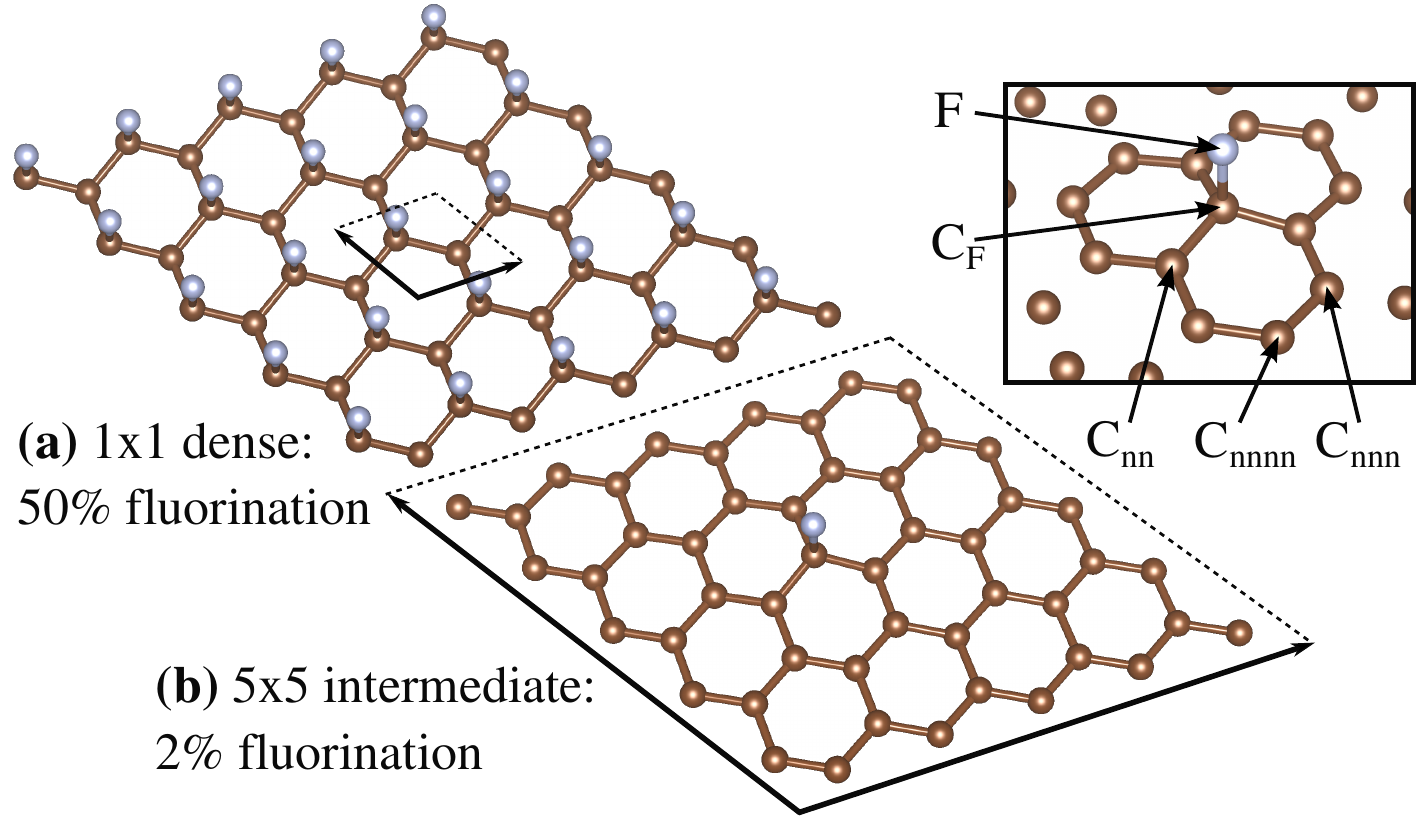}
 \caption{\label{fig:unit_cells}(Color online) Unit cells of the dense (50\%) and intermediate (2\%) fluorination limits. The lattice vectors are shown as black arrows and the unit cell is emphasized by the dashed lines. The inset shows the geometrical structure around a fluorine adatom, with labels for the adatom (F),
fluorinated carbon ($\rm C_F$), the nearest ($\rm C_{nn}$), next nearest ($\rm C_{nnn}$), and the next-next nearest
($\rm C_{nnnn}$) neighbors of the fluorinated carbon. }
\end{figure}

The unit cell of $\mathrm{C_2F}$ contains two carbon atoms and one fluorine which
chemisorbs preferentially in the top position, see Fig.~\ref{fig:unit_cells}. Our calculated Bader charges show a transfer of about $0.45e$ from graphene to fluorine. The charged fluorine adatoms repel each other,
stretching graphene whose lattice constant grows by about 4.1\% in comparison to the bare graphene lattice constant (a$_\text{Gr}=2.46$~\AA).
The graphene $\sigma$ bonds resist stretching, resulting in a regular $sp^3$-like corrugation; fluorinated carbon atoms ($\mathrm{C_F}$) are pulled out of the graphene plane by about $0.312$~\AA. In addition, our structural relaxation reveals that
$\mathrm{F}-\mathrm{C_F}$ bond length increases to 1.475~\AA\ in comparison to the typical carbon fluorine bond length \cite{Hagan2008:CSR} of 1.35~\AA. The local corrugation allows for $sp^3$-hybridization and hence for admixture of carbon $\pi$ and $\sigma$ states. This is expected to significantly enhance local spin-orbit coupling, in analogy to
hydrogenated graphene.\cite{Neto2009:PRL, Gmitra2013:PRL} On the other hand,
the fine structure of the fluorine atom shows a splitting of 50~meV ($\mathrm{^2P^0}$ spectroscopic term) \cite{Moore}---five times larger than in the carbon atom. This should further enhance SOC in fluorinated graphene due to the intrinsic fluorine contribution. In fact, as we show below, this contribution dominates the local enhancement of
spin-orbit coupling.

\begin{figure}[htp]
\includegraphics[width=0.98\columnwidth]{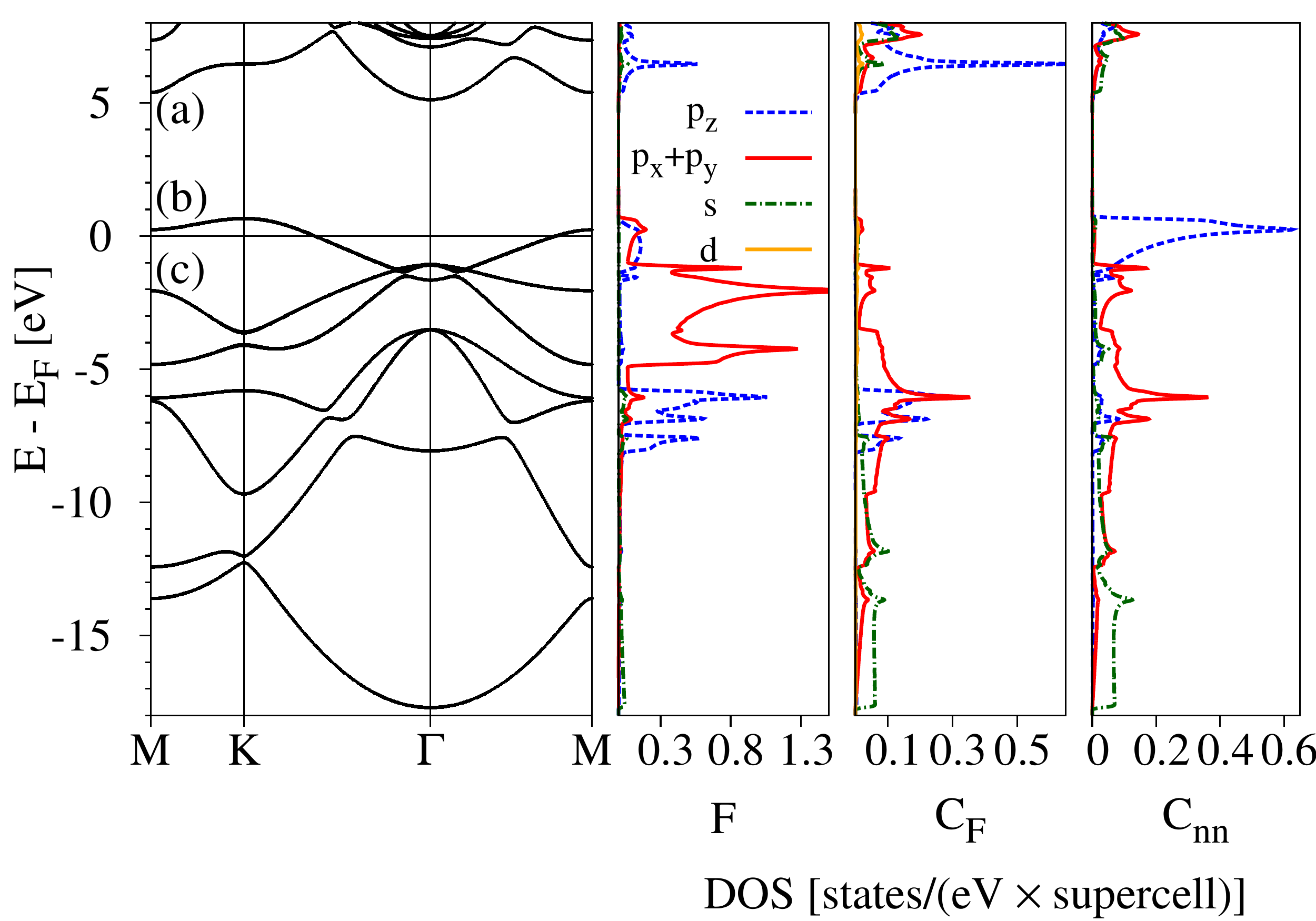}
\caption{\label{fig:GrF1so_banddos}(Color online) Calculated electronic properties of single-side semifluorinated graphene $\mathrm{C_2F}$ as obtained from nonmagnetic
\emph{ab initio} calculations. Shown is the electronic band structure (left) and the orbital resolved atomic density of states (right) for fluorine F, fluorinated carbon $\mathrm{C_F}$, and its nonfluorinated nearest neighbor $\mathrm{C_{nn}}$. Three relevant bands around the Fermi level are labeled by (a), (b), and (c), respectively. The band
(b), which crosses the Fermi level, is formed by the $p_z$ orbitals sitting on $\mathrm{C_{nn}}$.}
\end{figure}

In Fig.~\ref{fig:GrF1so_banddos} we show the calculated electronic band structure of
$\mathrm{C_2F}$. The fully occupied valence bands range in energies from $-23$ eV to $-1$ eV. There are
seven such bands which are occupied by 14 of the 15 valence electrons (the two carbons contribute
4 electrons each, and fluorine has 7 valence electrons). The band crossing the Fermi level, labeled as (b) in Fig.~\ref{fig:GrF1so_banddos},
is only partially occupied. The lowest valence band, at about $-23$ eV (not shown in the figure) stems from the
fluorine $2s$ orbitals; this band is only weakly dispersive.
In the energy region from $-18$ to $-8$ eV the band structure of $\mathrm{C_2F}$ is dominated by the almost intact
carbon $\sigma$ bonds, resembling the band structure of pristine graphene at those energies.
However, in the region from $-8$ to $-6$~eV we observe anticrossings in the band structure when moving away from the $\Gamma$ point. At these energies the orbital resolved DOS (see Fig.~\ref{fig:GrF1so_banddos}) on the fluorinated carbon $\mathrm{C_F}$ shows the presence of both $p_z$ and $p_x+p_y$ orbitals. This is the manifestation of the structural $sp^3$ hybridization---the $\mathrm{C_F}$ atom is pulled out of the plane, the associated
$p$ orbitals start to overlap and their interaction causes anticrossings.

Comparing the orbital resolved DOS for F and $\mathrm{C_F}$ atoms in Fig.~\ref{fig:GrF1so_banddos},
we see two energy windows $(-8, -6)$~eV and $(5.5, 7)$~eV in which both atoms contribute dominantly $p_z$ orbitals. These are the bonding and antibonding states
made of $p_z$ orbitals on $\mathrm{F}$ and $\mathrm{C_F}$.
On the other hand, the $p_z$ orbitals on the nearest neighbor carbon, $\mathrm{C_{nn}}$, form the band at the Fermi level that is partially occupied [indicated by (b) at Fig.~\ref{fig:GrF1so_banddos}]. This band is weakly dispersive (ca. 2~eV bandwidth) since the geometrically allowed interaction among different $\mathrm{C_{nn}}$ carbons is of the next-nearest-neighbor type. Finally, fluorine $p_x+p_y$ orbitals span the energy interval from $-5$~eV to $-1.5$~eV. These orbitals hybridize, due to their geo\-metry, only weakly with others. Contributions of different orbitals are indicated by the
labeled lines. We also note that there are small $d$ orbital contributions around the Fermi level which are most significant for the $\mathrm{C_F}$ atom.

\begin{figure}[htp]
\includegraphics[width=0.98\columnwidth]{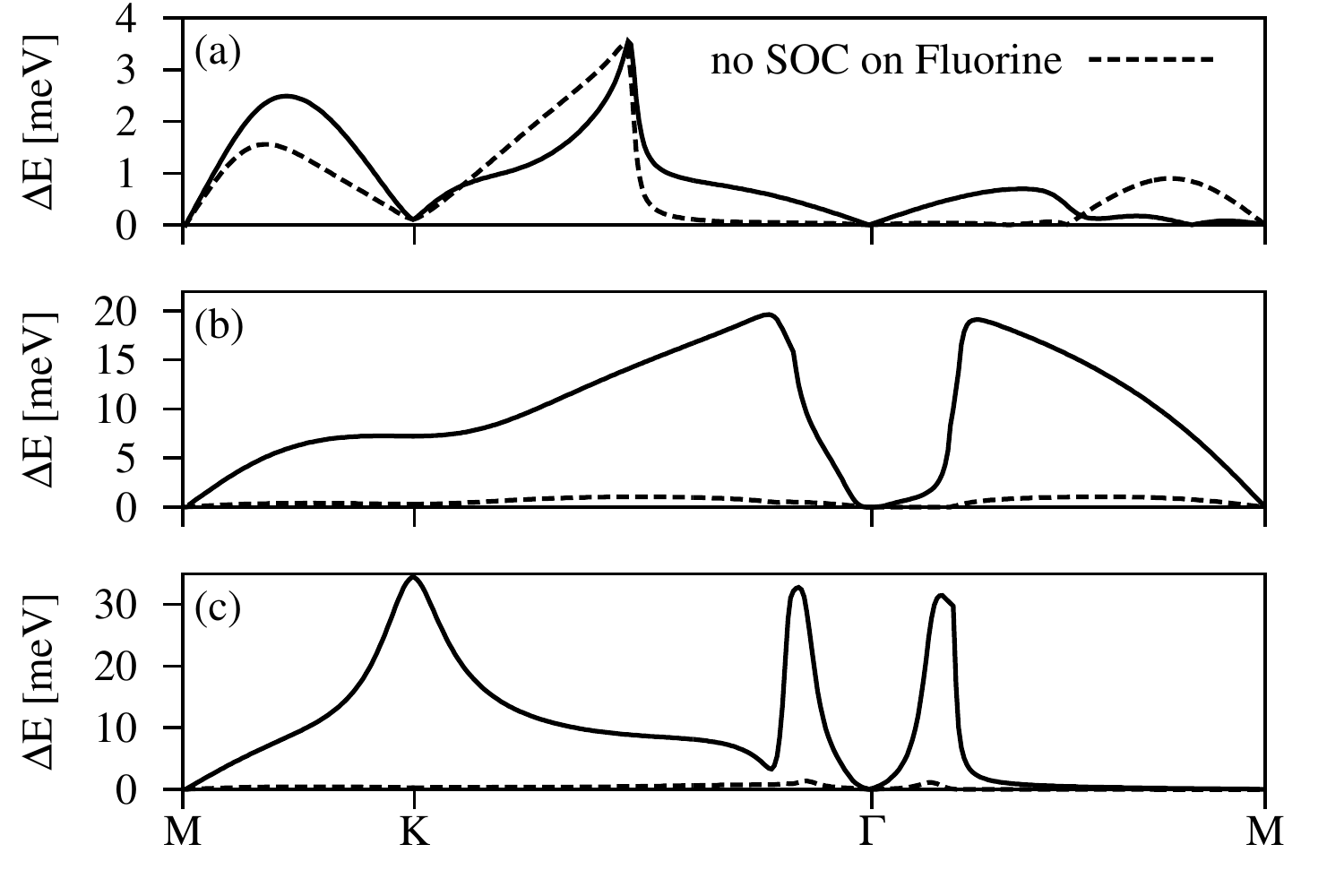}
\caption{\label{fig:GrF1so_splitting} Calculated band splittings due to SOC for the single-side semifluorinated graphene $\mathrm{C_2F}$.
Three figures---(a), (b), and (c)---correspond to SOC splittings of three relevant bands (a), (b), and (c) that are shown in
Fig.~\ref{fig:GrF1so_banddos}. Dashed lines correspond to the band splittings when turning off SOC on the fluorine adatom.}
\end{figure}

Spin-orbit splittings for the three bands around the Fermi level are shown in Fig.~\ref{fig:GrF1so_splitting}. The splittings vanish at the time reversal points $\Gamma$ and M and reach values up to 30~meV for band (c). The splitting is smallest for band (a), which is formed by the $p_z$ orbitals on the fluorinated carbon $\rm C_{F}$ and fluorine,
and has only a weak $p_x$ and $p_y$ component. As we turn off the spin-orbit coupling on $\rm F$ in our
first-principles calculations, this band still shows a significant splitting, in contrast to band (b) and band (c),
whose splitting drops to less than 1 meV. This demonstrates that band (a) has a significant part of its splitting due to $sp^3$ deformation alone.
On the other hand, bands (b) and (c) are formed to a large extent by fluorine $p_x$ and $p_y$ orbitals, whose atomic fine structure
is imprinted in the bands as the enhanced spin-orbit splitting.
The SOC splitting can be even more enhanced in the vicinity of the anticrossing points due to the more effective transfer of the spin-orbit coupling from the fluorine $p$ levels to the underlying graphene system. In the present case such anticrossing enhancements are observed around the $\Gamma$ point for the bands (b) and (c) (see the left panel of Fig.~\ref{fig:GrF1so_banddos} at about -1\,eV and the corresponding SOC splittings at Fig.~\ref{fig:GrF1so_splitting}), and also around the K point for the band (c) (see the left panel of Fig.~\ref{fig:GrF1so_banddos} at about -4\,eV and the corresponding SOC splitting at Fig.~\ref{fig:GrF1so_splitting}). The enhanced SOC splitting for the band (c) at the K point is attributed to the interaction with the band that is lying lower in energy at about $-4$~eV and which comprises mainly the fluorine $p_x$ and $p_y$ orbitals (see the orbital resolved electronic density at fluorine in Fig.~\ref{fig:GrF1so_banddos}).

\section{Dilute limit}

\subsection{DFT results for $10\times 10$ supercell}

To describe chemisorption of an isolated fluorine on graphene we need to consider a large enough supercell with a single fluorine
adatom to avoid interactions between periodic images of fluorine. We consider the fluorine atom on the top position, above a carbon atom, which has been reported as the energetically most favorable position.\cite{Sahin2011:PRB, Liu2012:JPCC} For hydrogenated graphene \cite{Gmitra2013:PRL} already a $5\times 5$ supercell is sufficient
to capture the essential features of the dilute limit. In the case of fluorinated graphene this supercell size is not enough as there is
still a significant overlap between the fluorine-derived states. We treat this intermediate case in the Appendix.
Here we focus on a $10 \times 10$ supercell (0.5\% of adatom coverage) as a representative of the dilute limit for fluorine adatoms.

Our structural relaxation shows that for the $10 \times 10$ supercell the $\mathrm{F}-\mathrm{C_F}$ bond length equals 1.607~\AA, the nearest-neighbor $\mathrm{C_F}-\mathrm{C_{nn}}$ bond length equals 1.469~\AA\ and the next-nearest-neighbor $\mathrm{C_{nnn}}-\mathrm{C_{nnn}}$ distance equals 2.488~\AA; see Fig.~\ref{fig:unit_cells}.
The fluorinated carbon $\mathrm{C_F}$ is pulled out of the graphene plane by about $0.423$~\AA, which is more than in
the dense limit ($0.312$~\AA).

\begin{figure}[htp]
\includegraphics[width=0.98\columnwidth]{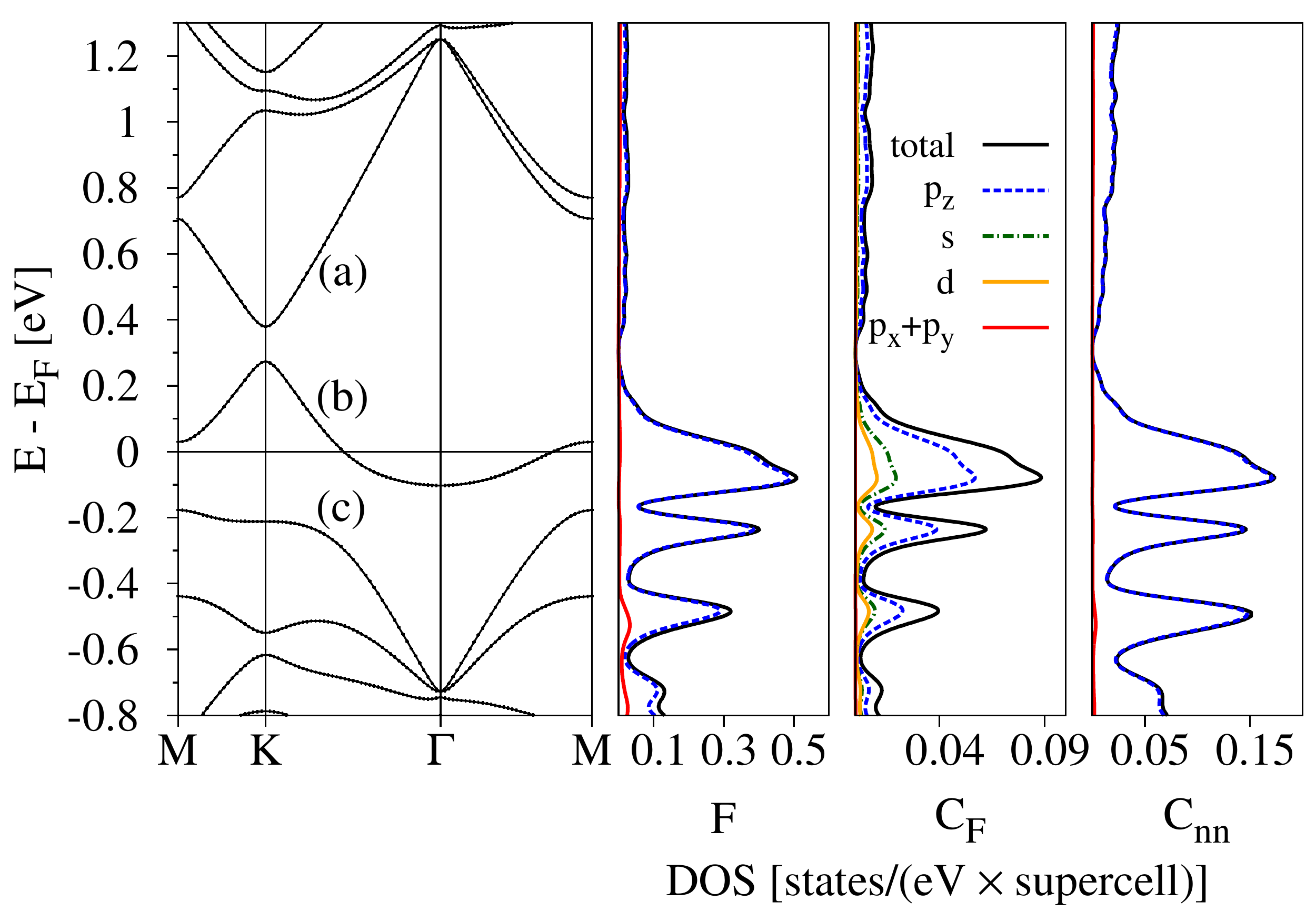}
\caption{\label{fig:GrF10_banddos}(Color online) Calculated electronic properties of fluor\-inated graphene in the dilute limit represented
by a $10\times 10$ supercell with a single fluorine adatom. Band structure (left) along high symmetry lines in first Brillouin
zone and orbital resolved atomic density of states (right) are shown. The labels correspond to conduction (a), midgap (b),
 and valence bands (c). Contribution of different orbitals is indicated by the
labeled lines.}
\end{figure}

In Fig.~\ref{fig:GrF10_banddos} we show the calculated electronic band structure and local DOS for selected atoms in the $10\times 10$
fluorinated graphene supercell. We again focus on three bands---(a), (b), and (c)---near the Fermi level which, in the following,
we call the conduction, midgap, and valence bands, respectively. The midgap band extends over the energy window from $-0.1$~eV to 0.3~eV and its bandwidth is by $0.2$~eV smaller than in the associated $5 \times 5$ system (see the Appendix). This implies that the Bloch modulation function of the midgap state becomes more localized and that the interaction among the fluorine adatoms is significantly weakened as compared to the more dense cases.
From the DOS it is apparent that the main contribution to the electron density of conduction, midgap, and valence bands comes from the $p_z$ orbitals on fluorine F and nearest neighbor $\mathrm{C_{nn}}$ carbon atoms. The orbital decomposition of the $\mathrm{C_{F}}$ DOS contains besides the large amount of $p_z$ also
$s$ and even $d$ orbitals. 
Like in the dense case also in the dilute system $d$ orbitals appear. Such contributions are also present in pure graphene where they play the crucial role for the intrinsic SOC. \cite{Gmitra2009:PRB} The situation here resembles the orbital decomposition in the $5\times 5$ system; see the Appendix.

The above orbital structure analysis suggests that a tight-binding model with only $p_z$ orbitals should be able to describe the
electronic structure in the vicinity of the Fermi level, i.e.,~the conduction, midgap, and valence bands
in the energy window from $-0.8$ to $0.8$~eV. Such a model is introduced in the following section.

\subsection{Effective Hamiltonian with spin-orbit coupling}
\subsubsection{Orbital part}

\begin{figure}
\includegraphics[width=\columnwidth]{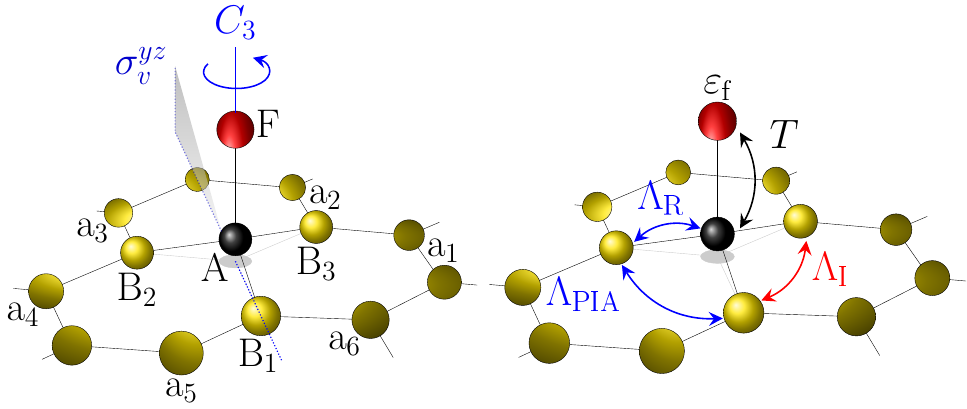}
\caption{\label{Fig:hop_par}(Color online) Left panel: mutual positions and adopted notation for all relevant atomic sites whose $p_z$ orbitals enter the model tight-binding Hamiltonian including local SOC. Shown are fluorine $\mathrm{F}$, fluorinated carbon $\mathrm{A}=\mathrm{C_F}$, its three nearest $\mathrm{B}_1,\mathrm{B}_2,\mathrm{B}_3$, and the six next-nearest $\mathrm{a}_1,\dots,\mathrm{a}_6$ neighbors.
The principal axis for $C_3$ rotations is defined by the perpendicular $\mathrm{F-A}$ bond, while $\sigma_v$-reflections are given by three $\mathrm{A-B}_i$ lines and the principal axis. Right panel: schematic representation of the dominant orbital and spin-orbital hoppings.}
\end{figure}

Employing local symmetries we derive and analyze an effective $p_z$ orbital tight-binding Hamiltonian including SOC relevant
in the vicinity of the fluorine adatom in the top position; see Fig.~\ref{Fig:hop_par}. While we focus on fluorine to obtain realistic
hopping parameters, the form of our Hamiltonian is transferable to other top-position adatoms on graphene.

Our first-principles analysis shows that the relevant states around the Fermi level originate from $p_z$ orbitals on carbon and fluorine atoms.
To describe the orbital part of the band structure (without spin effects) we employ a local hybridization Hamiltonian \cite{Robinson2008:PRL, Wehling2010:PRL, Kochan2014:PRL} which consists of an on-site energy $\varepsilon_{\text{f}}$ term on the fluorine adatom
and of the orbital hopping $T$ term between the $\mathrm{F}$ adatom and the fluorinated carbon $\mathrm{A}=\mathrm{C_F}$:
\begin{align}\label{Eq:horb}
 \mathcal{H}' = \varepsilon_{\text{f}}&\sum\limits_{\s} \hat{F}^{\dagger}_{\s}\hat{F}^{\phantom{\dagger}}_{\s} + T\sum\limits_{\s}\left(\hat{F}^{\dagger}_{\s}\hat{A}^{\phantom{\dagger}}_{\s} + \hat{A}_{\s}^{\dagger}\hat{F}^{\phantom{\dagger}}_{\s}\right)\,.
\end{align}
The rest is described by the standard nearest-neighbor hopping Hamiltonian for graphene,
\begin{equation}
\begin{aligned} \label{eq:H0}
\mathcal{H}_0 =&-t\sum\limits_{\langle i,j\rangle}\sum\limits_{\sigma} \left(\hat{c}_{i,\sigma}^\dagger \hat{c}_{j,\sigma}^{\phantom{\dagger}}+
\hat{c}_{j,\sigma}^\dagger \hat{c}_{i,\sigma}^{\phantom{\dagger}}\right)\\
&-t\sum\limits_{\mathrm{B}_j\in\mathrm{Cnn}}\sum\limits_{\sigma} \left(\hat{A}_{\sigma}^\dagger \hat{B}_{j,\sigma}^{\phantom{\dagger}}+
\hat{B}_{j,\sigma}^\dagger \hat{A}_{\sigma}^{\phantom{\dagger}}\right)\\
&-t\sum\limits_{\langle i,j\rangle}\sum\limits_{\sigma} \left(\hat{B}_{i,\sigma}^\dagger \hat{c}_{j,\sigma}^{\phantom{\dagger}}+
\hat{c}_{j,\sigma}^\dagger \hat{B}_{i,\sigma}^{\phantom{\dagger}}\right),
\end{aligned}
\end{equation}
with the orbital hopping $t=2.6$~eV; summation over $\langle i,j\rangle$ runs over all pairs of graphene nearest neighbors.
Operator $\hat{F}_{\s}^{\phantom{\dagger}}$ ($\hat{F}_{\s}^{\dagger}$) annihilates (creates) an electron with spin $\s$ in the atomic $p_{z}$ orbital on fluorine $\mathrm{F}$.
Similarly, $\hat{c}_{i\s}^{\phantom{\dagger}}$ ($\hat{c}_{i\s}^{\dagger}$) are the annihilation (creation) operators for $p_z$ orbitals
on graphene carbon atoms. We also introduce $\hat{A}_{\s}^{\phantom{\dagger}}$ ($\hat{A}_{\s}^{\dagger}$) and $\hat{B}_{i\s}^{\phantom{\dagger}}$ ($\hat{B}_{i\s}^{\dagger}$), $i=1,2,3$, as the annihilation (creation) operators on fluorinated
carbon site $\mathrm{A}$, as well as on its three nearest neighbors $\mathrm{B}_1,\mathrm{B}_2,\mathrm{B}_3$.
Our notation is illustrated in Fig.~\ref{Fig:hop_par}. Labels $A$ and $B$ derive from the corresponding sublattice; operators are typographically
distinguished to avoid confusion.

\begin{figure}
\includegraphics[width=0.85\columnwidth]{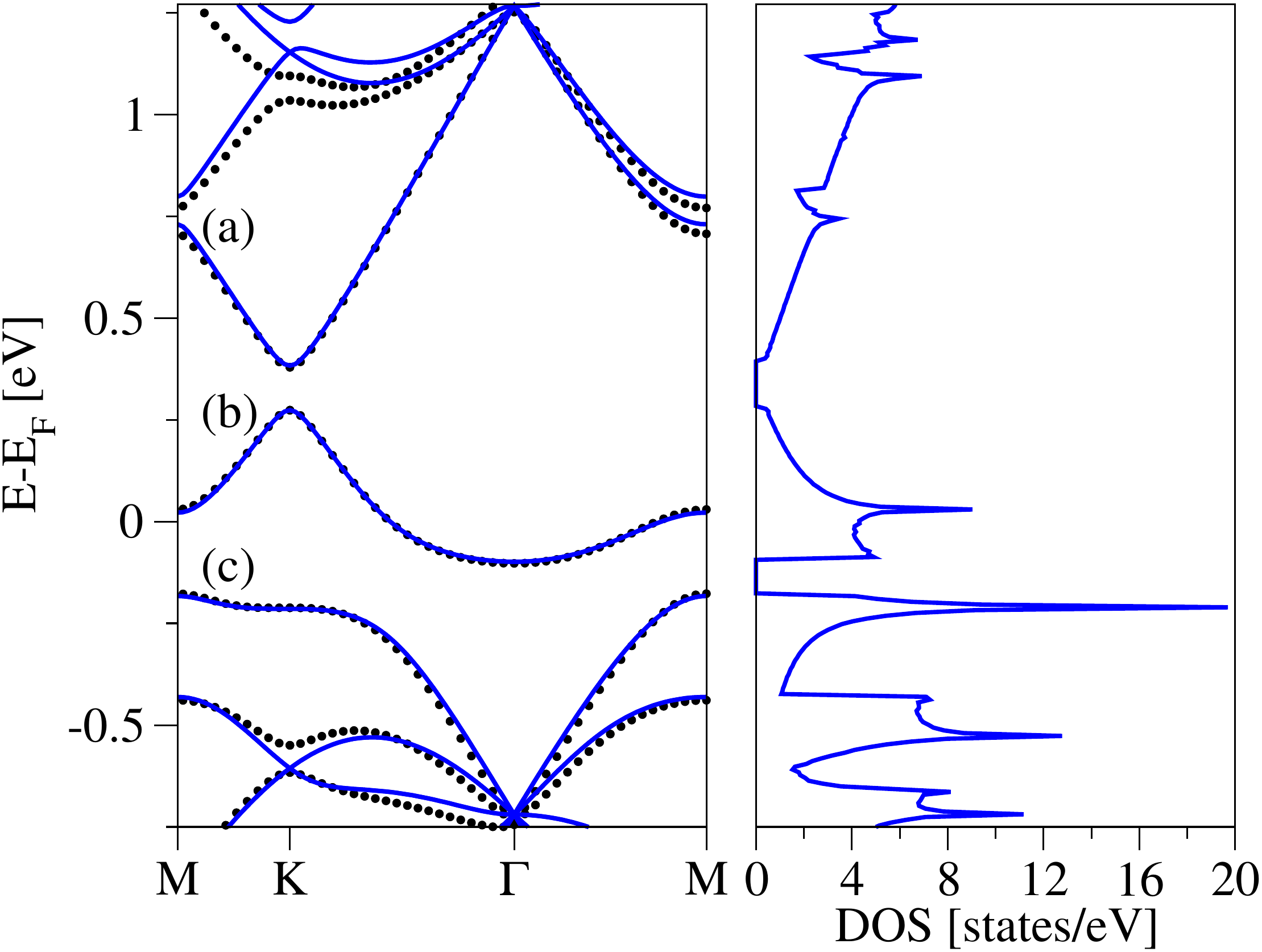}
\caption{\label{Fig:flgr10_banddos}(Color online) Left panel: electronic band structure of the dilute fluorinated graphene in $10\times 10$ supercell configuration; the first-principles data are presented by black dotted lines and the tight-binding model data by blue solid lines. The model calculation is based on the orbital Hamiltonian $\mathcal{H}_0+\mathcal{H}'$ with the parameters $T = 5.5$~eV and $\varepsilon_{\text{f}} = -2.2$~eV that are chosen by fitting the conduction (a), midgap (b), and valence band (c) along the M-K-$\Gamma$-M line, respectively.
Right panel: DOS per atom and spin corresponding to the tight-binding computed electronic band structure from the left panel.
}
\end{figure}

Hamiltonian $\mathcal{H}'$, Eq.~(\ref{Eq:horb}), is consistent with the structural $C_{3v}$ symmetry that emerges locally due to fluorine top-position chemisorption. We recall that the point group $C_{3v}$ is generated by $C_{3}$-rotations around the fluorine-carbon bond (principal axis) and $\sigma_v$---vertical reflections containing the principal axis and the $\mathrm{A}-\mathrm{B}_i$ bond. Using the full orbital Hamiltonian $\mathcal{H}_0+\mathcal{H}'$
we have fitted our first-principles band structure of the $10\times 10$ supercell.
To find the optimal values of two tight-binding parameters $T$ and $\varepsilon_{\text{f}}$ we have focused on conduction, midgap,
and valence bands around the Fermi level, see bands (a), (b), and (c) in Fig.~\ref{fig:GrF10_banddos}, as only those have a dominant $p_z$ character.
Minimization of the least-square differences between the first-principles and model-computed band structures gives $T = 5.5~\text{eV}$ and $\varepsilon_{\text{f}} = -2.2~\text{eV}$. The near perfect agreement with the first-principles data is shown in Fig.~\ref{Fig:flgr10_banddos}.
This figure also shows the model calculated DOS, using the triangle method (2d analog of the standard tetrahedron method).

\subsubsection{Spin-orbit part}

We now use symmetry arguments to derive the spin-orbit Hamiltonian $\mathcal{H}_{\mathrm{SO}}$ which gives the
fine splittings of the bands around the Fermi level. Again, only $p_z$ orbitals will be used in the description.
We identify non-zero matrix elements $\Braket{i,\s|H_{\text{SO}}|j,\s'}$ between atomic $p_z$ orbitals at sites $i$ and $j$ hosting spins $\s$ and $\s'$ using invariance of the microscopic spin-orbit Hamiltonian,
\begin{equation}
 H_{\text{SO}} = \frac{\hbar}{4m^2c^2}\,\bm{\hat{s}}\cdot\left(\bm{\nabla}\hat{V}\times\bm{\hat{p}}\right)\, ,
\end{equation}
under the relevant point-group symmetry and time reversal operations. Here, $m$ is the electron vacuum mass, $c$ the speed of light and $V$ the total electrostatic potential energy experienced by the electron. Symbols $\hat{p}$ and $\hat{s}$ are the operators of momentum and spin, respectively. As we wish to construct a minimal model, we limit the atomic orbitals $|i,\s\rangle$ basis of the SOC Hamiltonian to the immediate impurity region.
The lattice site index $i$ includes only fluorine $\mathrm{F}$,
fluorinated carbon $\mathrm{A}$, its three nearest $\mathrm{C_{nn}}=\{\mathrm{B}_1,\mathrm{B}_2,\mathrm{B}_3\}$, and its
six next-nearest neighbors $\mathrm{C_{nnn}}=\{\mathrm{a}_1,\dots,\mathrm{a}_6\}$; see Fig.~\ref{Fig:hop_par}.
Moreover, we take into account only those matrix elements $\Braket{i,\s|H_{\text{SO}}|j,\s'}$ for which $i$ and $j$ are spaced not further than up to the next-nearest neighbors.

Reduction of $D_{6h}$---the full point group symmetry of pristine graphene---to $C_{3v}$---the symmetry corresponding to the top-positioned adatom---induces several SOC mediated hoppings. Apart from the usual intrinsic and Rashba hoppings, which are found in gated graphene or in graphene
on a substrate,\cite{Konschuh2010:PRB} there are more terms allowed. In what follows we summarize and discuss all the allowed
matrix elements $\Braket{i,\s|H_{\text{SO}}|j,\s'}$ in the specified impurity region.
Since time-reversal symmetry enables us to classify which parameters are real, imaginary, or generally complex,
we chose the representative SOC parameters $\Lambda$'s shown below to be real.
The result is as follows:
\begin{subequations}\label{Eq:SOC-couplings}
\begin{align}
\LIA&=\tfrac{3\sqrt{3}}{\ii}\Braket{\mathrm{A},\ua|H_{\text{SO}}|\mathrm{a}_1,\ua},\label{Eq:SOC-couplingsa}\\
\LIB&=-\tfrac{3\sqrt{3}}{\ii}\Braket{\mathrm{B}_2,\ua|H_{\text{SO}}|\mathrm{B}_3,\ua},\label{Eq:SOC-couplingsb}\\
\LBR&=\tfrac{3}{2\ii}\Braket{\mathrm{A},\ua|H_{\text{SO}}|\mathrm{B}_1,\da},\label{Eq:SOC-couplingsc}\\
\LPA+\ii\tilde{\Lambda}^\mathrm{A}_{\mathrm{PIA}}&=\tfrac{3}{2}\Braket{\mathrm{A},\ua|H_{\text{SO}}|\mathrm{a}_1,\da},\label{Eq:SOC-couplingsd}\\
\LPB&=-\tfrac{3}{2}\Braket{\mathrm{B}_2,\ua|H_{\text{SO}}|\mathrm{B}_3,\da},\label{Eq:SOC-couplingse}\\
\LFC&=\tfrac{3}{2\ii}\Braket{F,\ua|H_{\text{SO}}|\mathrm{B}_1,\da}.\label{Eq:SOC-couplingsf}
\end{align}
\end{subequations}
Before we discuss the resulting hoppings, we make two technical comments.
First, to obtain the actual phase factors for the matrix elements that correspond to similar atomic orbital configurations, e.g.,
for $\Braket{\mathrm{A},\ua|H_{\text{SO}}|\mathrm{a}_2,\da}$, one needs to employ appropriate symmetry operations.
This is already accounted for in the SOC Hamiltonian $\mathcal{H}_{\mathrm{SO}}$ provided below.
Second, the numerical prefactors are chosen such that the expansion of the SOC Hamiltonian near the K point
is numerically prefactor-free, when considering a system coated by adatoms periodically (forming a supercell).

Let us discuss the first two terms in Eqs.~(\ref{Eq:SOC-couplingsa}) and (\ref{Eq:SOC-couplingsb}). Those are sublattice-resolved local
intrinsic SOCs: $\LIA$ mediates spin-conserving hopping between fluorinated carbon $\mathrm{A}$ and its six next-nearest neighbors $\mathrm{C_{nnn}}=\{\mathrm{a}_1,\dots,\mathrm{a}_6\}$, while $\LIB$ does the same for the nearest carbon atoms $\mathrm{C_{nn}}=\{\mathrm{B}_1,\mathrm{B}_2,\mathrm{B}_3\}$; refer to Fig.~\ref{Fig:hop_par} for the labeling.
Equation~(\ref{Eq:SOC-couplingsc}) describes the local Rashba coupling that mediates the spin-flip hopping between fluorinated carbon $\mathrm{A}$ and its three nearest neighbors $\mathrm{C_{nn}}$.
The origin of the Rashba coupling, and hence of the fixed orientation of the perpendicular $z$ axis stems from a local dipolar electric field that appears due to charge redistribution caused by the fluorine chemisorption.
The induced dipolar field can also affect the spin-flip coupling between the nearest and next-nearest neighbor sites, $\mathrm{C_{nn}}$ and $\mathrm{C_{nnn}}$, but we do not consider terms like $\Braket{\mathrm{B}_1,\ua|H_{\text{SO}}|\mathrm{a}_6,\da}$ in what follows.

The three terms in Eqs.~(\ref{Eq:SOC-couplingsd})$-$(\ref{Eq:SOC-couplingsf}) are specific for systems with $C_{3v}$ structural symmetry.
They are describing spin-flip hoppings which connect the {\it next}-nearest neighbor sites.
The first two terms in Eqs.~(\ref{Eq:SOC-couplingsd}) and (\ref{Eq:SOC-couplingse}) are sublattice-resolved PIA-SOC terms, introduced already for hydrogenated
graphene \cite{Gmitra2013:PRL} and silicene.\cite{Geissler2013:NJP}
The acronym PIA stands for \emph{pseudospin-inversion-asymmetry} induced SOC: compared to pristine graphene,
sublattices A and B are, close to the impurity site, not equivalent. The generally complex matrix element
$\Braket{\mathrm{A},\ua|H_{\text{SO}}|\mathrm{a}_i,\da}$ connects fluorinated carbon A with its six next-nearest neighbors $\mathrm{C_{nnn}}$.
We have checked that the imaginary part of this matrix element plays only a minor role in the energy band splittings, so we set $\LPA\simeq\tfrac{3}{2}\Braket{\mathrm{A},\ua|H_{\text{SO}}|\mathrm{a}_1,\da}$ to be real in what follows.
(In the dense limit, corresponding to the fully fluorinated sublattice, this approximation becomes exact, since
the restored translational symmetry prohibits the imaginary part $\tilde{\Lambda}^\mathrm{A}_{\mathrm{PIA}}$.\cite{Gmitra2013:PRL})

Similarly, the real SOC $\LPB$ mediates spin-flip hoppings among three nearest neighbors $\mathrm{B}_1$, $\mathrm{B}_2$, and $\mathrm{B}_3$.
Our analysis shows that this term is crucial for explaining the fine band structure splittings due to adatom induced SOC.
Finally, parameter $\LFC$ describes allowed spin-flip hoppings among the fluorine F and three $\mathrm{C_{nn}}$ carbon atoms $\mathrm{B}_1$, $\mathrm{B}_2$, and $\mathrm{B}_3$. In our fitting procedure we find that for fluorine adatoms this term is negligible.

We now give a minimal SOC Hamiltonian for a top-position chemisorbed adatom:
\begin{align}
 \mathcal{H}_{\text{SO}} =& \frac{\text{i}\LIA}{3\sqrt{3}}\sum\limits_{c_j \in \rm{C_{nnn}}}\sum\limits_{\s}\left[\hat{A}^{\dagger}_{\s} \nu_{ij} \left(\hat{s}_{z}\right)_{\s\s}\hat{c}_{j,\s}+ \mathrm{h.c.}\right] \nonumber\\
 +&\frac{\text{i}\LIB}{3\sqrt{3}}\sum\limits_{\llangle i,j\rrangle}\sum\limits_{\s}\hat{B}^{\dagger}_{i,\s} \nu_{ij} \left(\hat{s}_{z}\right)_{\s\s}\hat{B}_{j,\s} \nonumber\\
 +&\frac{2\text{i}\LBR}{3}\sum\limits_{\mathrm{B}_j \in \rm{C_{nn}}}\sum\limits_{\s\neq\s'} \left[\hat{A}^{\dagger}_{\s}\left(\bm{\hat{s}}\times\bm{d}_{\mathrm{A}j}\right)_{z,\s\s'}\hat{B}_{j,\s'} + \mathrm{h.c.}\right]\nonumber\\
 +&\frac{2\text{i}\LPA}{3}\sum\limits_{c_j \in \rm{C_{nnn}}}\sum\limits_{\s\neq\s'} \left[\hat{A}^{\dagger}_{\s}\left(\bm{\hat{s}}\times\bm{D}_{\mathrm{A}j}\right)_{z,\s\s'}\hat{c}_{j,\s'}+ \mathrm{h.c.}\right]\nonumber\\
 +&\frac{2\text{i}\LPB}{3}\sum\limits_{\llangle i,j\rrangle}\sum\limits_{\s\neq\s'} \hat{B}^{\dagger}_{i,\s}\left(\bm{\hat{s}}\times\bm{D}_{ji}\right)_{z,\s\s'}\hat{B}_{j,\s'}\nonumber\\
 +&\frac{2\text{i}\LFC}{3}\sum\limits_{\mathrm{B}_j \in \rm{C_{nn}}}\sum\limits_{\s\neq\s'} \left[\hat{F}^{\dagger}_{\s}\left(\bm{\hat{s}}\times\bm{d}_{\mathrm{F}j}\right)_{z,\s\s'}\hat{B}_{j,\s'} + \mathrm{h.c.}\right] \nonumber\\
 +&\frac{\text{i}\lI}{3\sqrt{3}}{\sum\limits_{\llangle i,j\rrangle}}'\sum\limits_{\s}\hat{c}^{\dagger}_{i,\s} \nu_{ij} \left(\hat{s}_{z}\right)_{\s\s}\hat{c}_{j,\s}\,.\label{Eq:SOC_Hamiltonian}
\end{align}
The creation and annihilation operators are defined in the discussion after Eq.~(\ref{eq:H0}).
Symbol $\bm{\hat{s}}$ represents the array of Pauli matrices acting on the spin space; the sign factor $\nu_{ij}$
equals $+1$ if the next-nearest neighbor hopping path $j\rightarrow k\rightarrow i$ via a common neighbor $k$
is counterclockwise ($-1$ for clockwise).
Vectors $\bm{d}_{ij}$ and $\bm{D}_{ij}$ are unit vectors in the $xy$ plane (perpendicular to the $\mathrm{F-A}$ bond)
pointing from the projected site $j$ to $i$.
The last term in Eq.~(\ref{Eq:SOC_Hamiltonian}) is the intrinsic SOC of pristine graphene for which $\lambda_{\mathrm{I}}=12~\mu$eV, see Ref.~\onlinecite{Gmitra2009:PRB}. We have implemented this term for all next-nearest neighbors not participating in SOC hoppings with the coupling constants $\LIA$ and $\LIB$; this fact is indicated by the prime at the corresponding summation symbol.

\begin{figure}
\includegraphics[width=0.85\columnwidth]{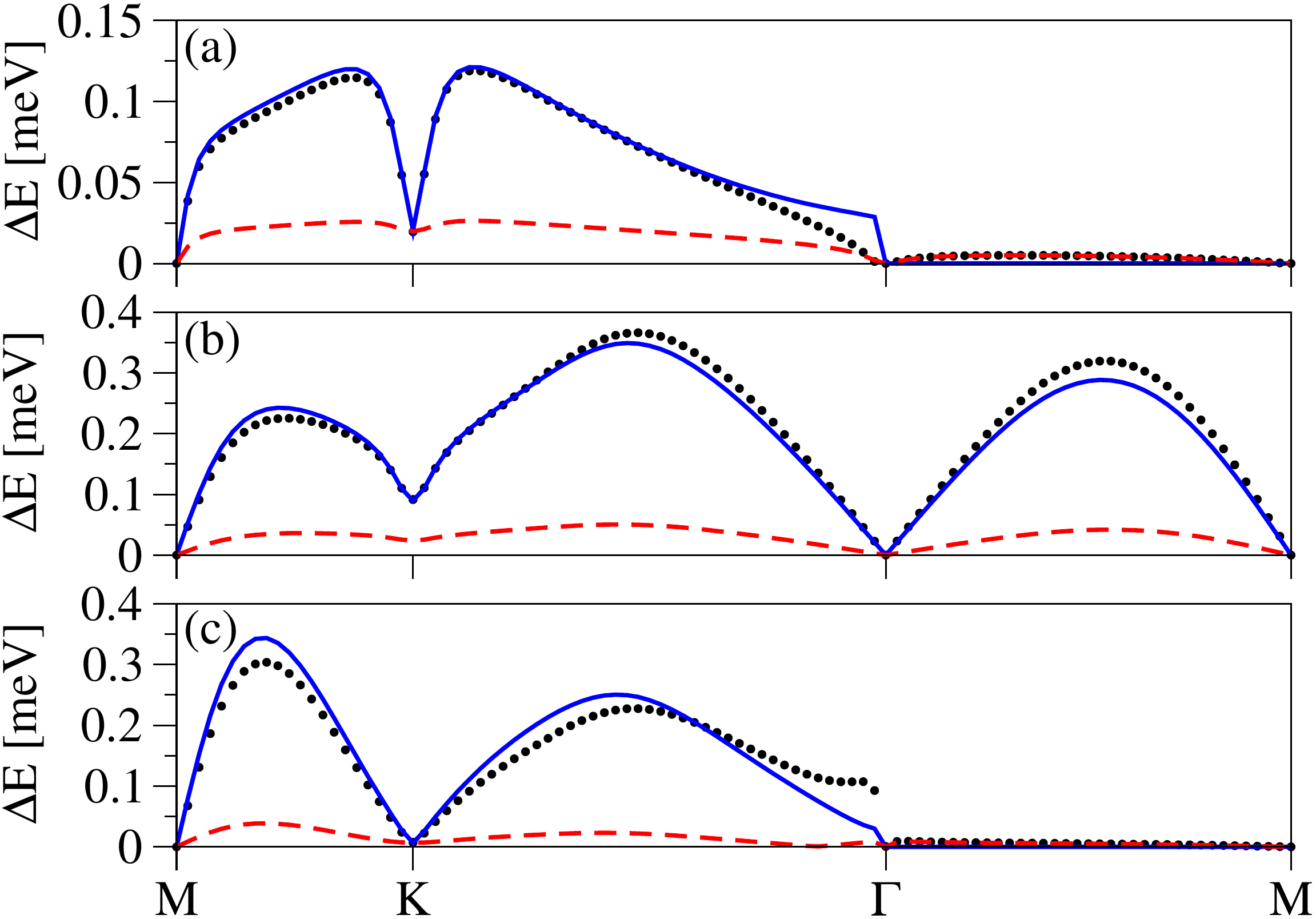}
\caption{\label{fig:SOC_splittings_model_vs_DFT}(Color online)
Spin-orbit splittings along the M-K-$\Gamma$-M line for the conduction (a), midgap (b), and valence bands (c); refer also
to Fig.~\ref{Fig:flgr10_banddos}.
First-principles data (dotted) are excellently reproduced by the phenomenological tight-binding model (solid) given by the Hamiltonian
$\mathcal{H}_0+\mathcal{H}'+\mathcal{H}_{\mathrm{SO}}$ from the text.
The following orbital and SOC parameters were used in the tight-binding model as the best fits: $T = 5.5$~eV, $\varepsilon_{\text{f}} = -2.2$~eV, $\Lambda^{\text{B}}_{\text{I}} = 3.3$~meV, $\Lambda^{\text{B}}_{\text{PIA}} = 7.3$~meV, and $\Lambda_{\text{R}} = 11.2$~meV.
Turning off the intra-atomic SOC of the fluorine adatom in the first-principles calculations reduces the spin-orbit splitting significantly (dashed line).
This residual spin-orbit splitting can be attributed to the structural local $sp^3$ distortion caused by fluorination.
}
\end{figure}

\subsubsection{Fits to first-principles results}

The fit of our tight-binding model to the first-principles calculations of the $10 \times 10$ supercell of fluorinated graphene is given
in Fig.~\ref{fig:SOC_splittings_model_vs_DFT}. Spin-orbit splittings of bands (a), (b), and (c) from Fig.~\ref{fig:GrF10_banddos}
are shown. The splittings reach maxima of 0.1, 0.35, and 0.3 meV for the three (a), (b), and (c) bands, respectively.
Keeping the orbital parameters fixed we focused on reproducing SOC induced splittings employing the full model Hamiltonian $\mathcal{H}_0+\mathcal{H}'+\mathcal{H}_{\mathrm{SO}}$.
We explored various combinations of $\Lambda$'s that enter Eq.~(\ref{Eq:SOC_Hamiltonian}) to generate a minimum robust set of parameters capable to explain the observed SOC splittings.
Minimizing the sum of least-square differences for SOC splittings of conduction, midgap, and valence bands along the
full M-K-$\Gamma$-M line, we have obtained the minimal SOC basis with $\LIB=3.3$~meV, $\LPB=7.3$~meV, and $\LBR=11.2$~meV only; see Fig.~\ref{fig:SOC_splittings_model_vs_DFT}. This indicates that all remaining group-theory allowed SOC parameters
in Hamiltonian $\mathcal{H}_{\rm{SO}}$, Eq.~(\ref{Eq:SOC_Hamiltonian}), can be safely omitted.
For the valence and conduction bands near the $\Gamma$ point we observe marked discrepancies between the
tight-binding model and first-principles calculations.
The reason is the admixture of $2s$ and $3d$ orbitals that contribute to DOS as much as $2p_z$ orbitals near $\Gamma$, i.e.,~at energies below
$-0.6$~eV for the valence band and above $0.9$~eV for the conduction band; see Fig.~\ref{fig:GrF10_banddos}.
Since our symmetry inspired effective SOC Hamiltonian accounts for $p_z$ orbitals, discrepancies such as those around $\Gamma$ are
to be expected. Nevertheless, it is rather remarkable that with only three SOC parameters we are able to near perfectly match
all the characteristic features accompanying SOC splittings along the {\it whole} M-K-$\Gamma$-M line for all three bands around the Fermi level.
Such a close agreement with first-principles results is to some extent fortuitous, given by the almost sole $p_z$ character of the
bands around the Fermi level. At the same time it gives us confidence in our minimal phenomenological orbital and SOC hopping model,
whose graphical representation is shown in Fig.~\ref{Fig:hop_par}.

In addition to the $10\times 10$ supercell, we have also calculated a smaller, $7\times 7$, supercell from first principles.
The best-fit tight-binding model parameters in this case are $T = 6.1$~eV, $\varepsilon_{\text{f}} = -3.3$~eV, and $\LIB = 3.2$~meV, $\LPB = 7.9$~meV, and $\LBR = 11.3$~meV.
These values are very close to those of the $10\times 10$ supercell, further evidencing the robustness and consistency of our minimal tight-binding model. The spin-orbit parameters seem to be less sensitive to the supercell size than the orbital ones. Both supercell results are summarized in Table~\ref{Tab:7x7-vs-10x10}.
The intermediate case of a $5\times 5$ supercell, which we find to be insufficient to describe the
dilute limit, and so unsuitable for our tight-binding analysis here, is treated in the Appendix.

\begin{table}
  \begin{ruledtabular}
    \begin{tabular}{cccccc}
    $n\times n$ & $T$\,[eV] & $\varepsilon_\text{f}$\,[eV] & $\Lambda^{\text{B}}_{\text{I}}$\,[meV] & $\Lambda^{\text{B}}_{\text{PIA}}$\,[meV] & $\Lambda_{\text{R}}$\,[meV] \\\hline\\
    $7\times 7$ & 6.1 & -3.3 & 3.2 & 7.9 & 11.3\\
    $10\times 10$ & 5.5 & -2.2 & 3.3 & 7.3 & 11.2
    \end{tabular}
  \end{ruledtabular}
  \caption{\label{Tab:7x7-vs-10x10}Orbital and SOC tight-binding parameters that fit the electronic band structure of fluorinated graphene in $7\times 7$ and $10\times 10$ supercell configurations, respectively. The corresponding values are relatively close for both supercells, confirming the reliability and robustness of our phenomenological tight-binding model. The rest of the group-theory allowed SOC parameters, as defined by Eqs.~(\ref{Eq:SOC-couplings}), can be set to zero.}
\end{table}

As in the dense limit, there are two principal causes of the enhanced spin-orbit coupling due to fluorine adatom: $sp^3$ hybridization of
carbon orbitals (with spin-orbit from the $\sigma$ bonds) and the native spin-orbit coupling of fluorine. In order to separate and quantify these two contributions,
we turned off the intra-atomic SOC on fluorine in the first-principles calculation; see Fig.~\ref{fig:SOC_splittings_model_vs_DFT}.
The splitting is significantly reduced, by an order of magnitude. We conclude that $sp^3$ distortion gives spin-orbit splitting in
magnitude a decade smaller than what is induced by the native spin-orbit coupling of graphene. The giant enhancement of spin-orbit
coupling in graphene due to fluorine adatoms comes almost solely from the spin-orbit coupling of fluorine and hybridization of fluorine
$p$ orbitals with those of carbon. To understand the microscopic mechanism of this SOC transfer one should build a multiorbital tight-binding model with all relevant orbitals on $\mathrm{F}$, $\mathrm{A}$, and $\mathrm{B}_1,\mathrm{B}_2,\mathrm{B}_3$ sites and then down-fold the Hamiltonian matrix to the $p_z$ orbital sector, as was done for graphene.\cite{Konschuh2010:PRB, Kormanyos2013:PRB}
Such an analysis is beyond the scope of the present paper which aims at presenting an effective single-orbital
model, given by Hamiltonian $\mathcal{H}_{\mathrm{SO}}$, Eq.~(\ref{Eq:SOC_Hamiltonian}),
that can be used for transport and spin relaxation studies.

In contrast, in hydrogenated graphene the microscopic physics behind the giant enhancement of spin-orbit coupling
is $sp^3$ bonding, as shown by first-principles calculations.\cite{Gmitra2013:PRL} The spin-orbit coupling comes
from the carbon $\sigma$ bonds, which are split by about 10 meV at $\Gamma$. Part of this splitting is transferred
to the $\pi$ band upon $sp^3$ hybridization as a hydrogen is added on top of a carbon atom.
In fact, spin-orbit splittings induced by hydrogen are similar in magnitude as what is represented by
the dashed lines in Fig.~\ref{fig:SOC_splittings_model_vs_DFT}: compare with Ref.~\onlinecite{Gmitra2013:PRL}.
The effective SOC Hamiltonian $\mathcal{H}_{\mathrm{SO}}$, Eq.~(\ref{Eq:SOC_Hamiltonian}), can be also
employed in the hydrogenated case, for which the most dominant SO couplings are $\LIA=-0.21$~meV, $\LPB=-0.77$~meV, and $\LBR=0.33$~meV. Comparing spin-orbit coupling in pristine, hydrogenated, and fluorinated graphene, the magnitude grows roughly from $10~\mu\mathrm{eV}$, $1~\mathrm{meV}$, to $10~\mathrm{meV}$, respectively, reflecting
different microscopic mechanisms behind the coupling in these systems.

\subsection{``Is fluorine on graphene a resonant scatterer?''}

\begin{figure}
\includegraphics[width=0.85\columnwidth]{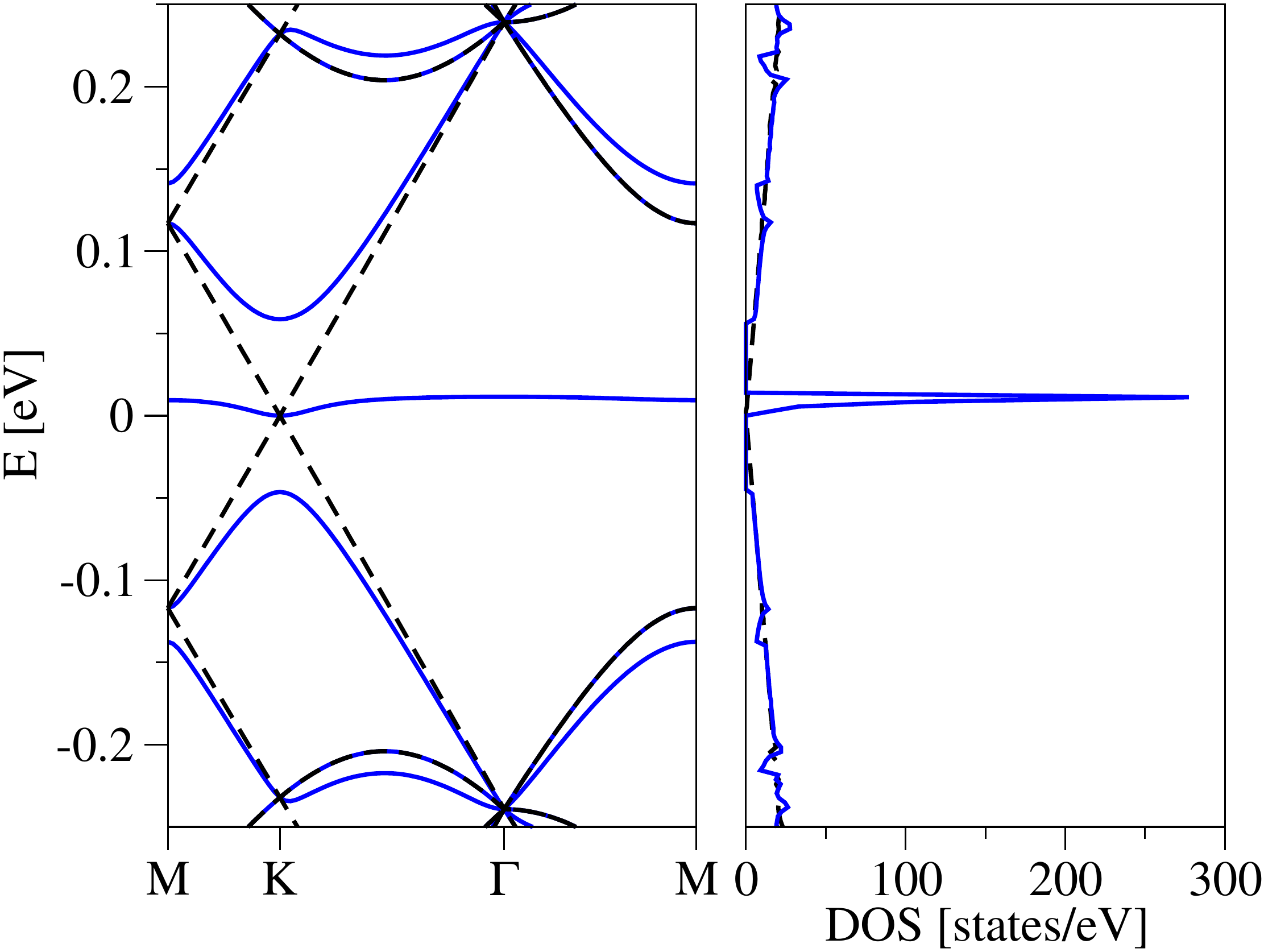}
\caption{\label{Fig:hgr40_banddos}(Color online) Extrapolated band structure and DOS of pristine (dashed) and hydrogenated (solid) graphene for a $40\times 40$ supercell. The left panel shows tight-binding model-calculated band structures along M-K-$\Gamma$-M lines. The right panel shows the corresponding DOS.
Tight-binding orbital parameters for hydrogenated graphene, $T_{\text{h}}=7.5$~eV and $\varepsilon_{\text{h}} = 0.16$~eV, are taken from Ref.~\onlinecite{Gmitra2013:PRL}.
The band structure of hydrogenated graphene shows a narrow resonant peak in the vicinity of the Dirac point.
}
\end{figure}

Hydrogen chemisorbed on graphene acts as a resonant scatterer,\cite{Wehling2010:PRL,Kochan2014:PRL} giving
a narrow pronounced peak in the DOS close to the Dirac point. In Fig.~\ref{Fig:hgr40_banddos} we plot the
tight-binding bands and DOS for hydrogenated graphene using a $40\times40$ supercell with a single hydrogen in
the top position. The midgap band, formed of $p_z$ orbitals on the nearest-neighbor $\rm C_{nn}$ carbon
is very narrow, developing the resonance peak seen in the DOS. As the supercell size grows, the valence and conduction bands merge towards the Dirac cone structure of pristine graphene, leaving the flat midgap band almost intact.
Such resonances are predicted to have profound effects on spin relaxation \cite{Kochan2014:PRL} and spin transport.\cite{Ferreira2014:PRL}

\begin{figure}
\includegraphics[width=0.85\columnwidth]{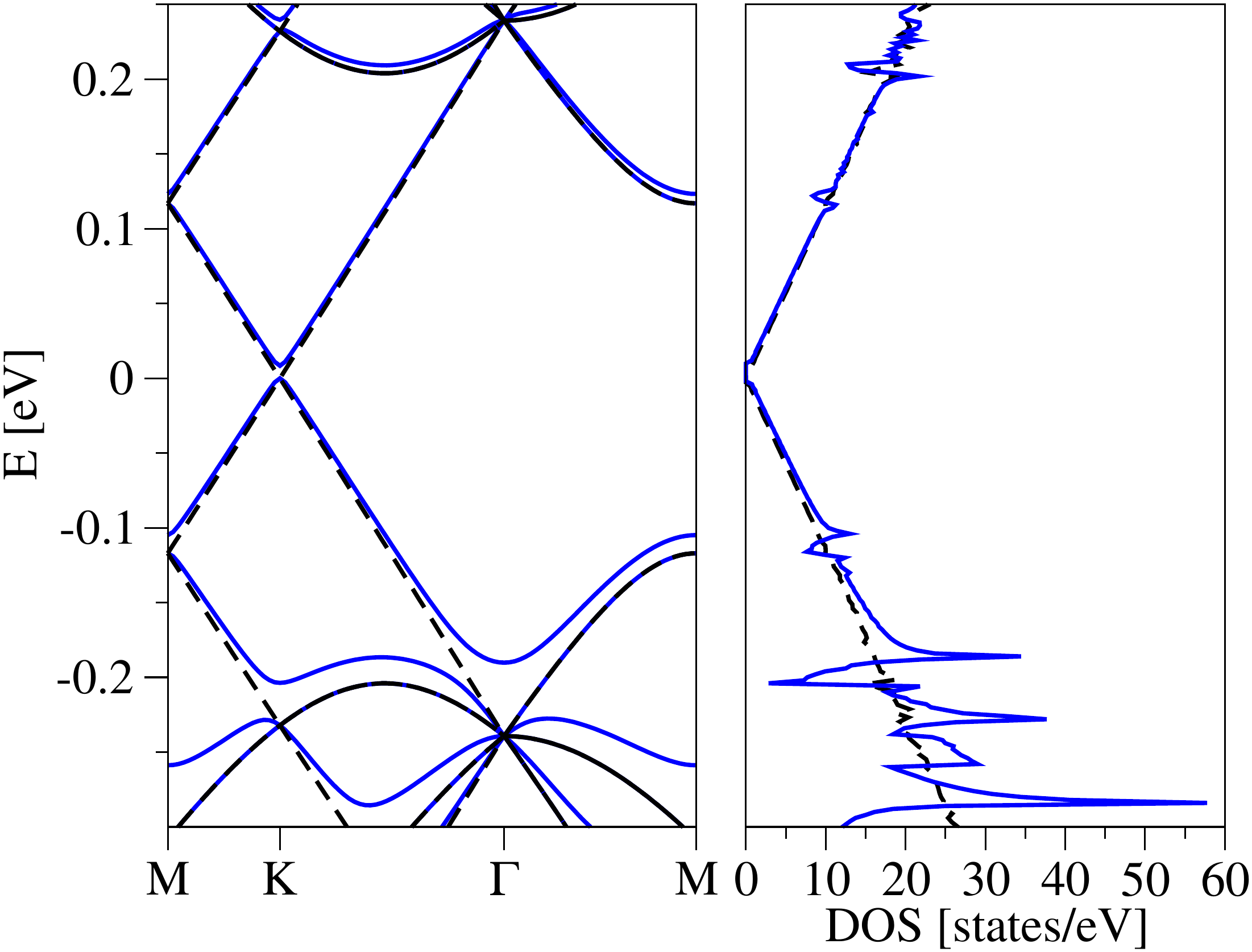}
\caption{\label{Fig:flgr40_banddos}(Color online) Extrapolated band structure and DOS for the pristine (dashed) and fluorinated (solid) graphene for a $40\times 40$ supercell configuration.
The left panel shows model-calculated band structures along M-K-$\Gamma$-M and the right panel the corresponding DOS. Characteristic changes in DOS
for the dilute fluorinated graphene become obvious below $-0.1$~eV. Tight-binding orbital parameters used in the model calculations, $T=5.5$~eV and $\varepsilon_{\text{f}}=-2.2$~eV, are best-fits to DFT results for $10\times 10$ supercell configuration.
}
\end{figure}

Fluorinated graphene in the dilute limit looks qualitatively different from hydrogenated graphene. The calculated
tight-binding band structure of a $40\times 40$ supercell is shown in Fig.~\ref{Fig:flgr40_banddos}.
We have used the same parameters as for the $10\times 10$ supercell in Table \ref{Tab:7x7-vs-10x10}, which we
believe are representative for the dilute limit. The valence and conduction bands look similar to the corresponding
bands of pristine graphene. The Dirac structure is almost intact! What used to be the midgap band (b) merges
together with the conduction band (a), creating close to the K point a superimposed band structure with linear dispersion.
The linear behavior is clearly seen at small energies in the DOS for which the gap at K is gradually decreasing.
No resonant level is observed close to the Dirac point. However, there are significant changes in the Dirac cone structure
occurring at energies below $-0.1$~eV, as seen in Fig.~\ref{Fig:flgr40_banddos}. Here, DOS shows a multiple-peak structure
due to anticrossings of the bands.

The question arises of what would be the ultimate limit of a single fluorine adatom on an infinite graphene sheet.
The above $40\times 40$ supercell calculations indicate a broad resonance peaked somewhere below $-0.2$~eV.
Fortunately, the orbital-hopping part of our Hamiltonian is analytically solvable in the single-adatom limit,
and we can calculate the change $\Delta\nu(E)$ of the DOS due to a single fluorine adatom.
Considering a very low concentration $\eta$ (just a prefactor) of adatoms,
a typical change in the unperturbed DOS per atom and spin---ignoring multiple-scattering interference among different adatom scatterers---would be given by $\eta\,\Delta\nu(E)$.
The tight-binding Hamiltonian (\ref{Eq:horb}) can be effectively downfolded, eliminating the fluorine $p_z$ orbitals by means of the L\"{o}wdin-Schrieffer-Wolff transformation.
As a result $\mathcal{H}'$ produces an energy dependent delta-like interaction
$\mathcal{H}'_{\mathrm{fold}}(E)$ localized at the fluorinated carbon $\rm C_F$,
\begin{equation}
\mathcal{H}'_{\mathrm{fold}}(E)=\sum\limits_\sigma\,\alpha(E)\,\hat{A}^\dagger_\sigma\,\hat{A}^{\phantom{\dagger}}_\sigma\,,
\ \mbox{where}\ \alpha(E)=\frac{T^2}{E-\epsilon_{\mathrm{f}}}\,.
\end{equation}
Following Ref.~\onlinecite{Hewson1993}, we obtain,
\begin{equation}\label{Eq:DOS-change}
\Delta\nu(E)=\frac{1}{\pi}\,\mathrm{Im}\Bigl[\frac{\alpha(E)}{1-\alpha(E)\,G_0(E)}\,\frac{\partial}{\partial E}\, G_0(E)\Bigr]\,.
\end{equation}
Here, $G_0(E)$ is the Green's function per atom and spin for the unperturbed pristine graphene, which, within the energy range from $-0.5$~eV up to $0.5$~eV, can be very well approximated by
\begin{equation}
G_0(E)\simeq\frac{E}{D^2}\Bigl[\ln{\Bigl|\frac{E^2}{D^2-E^2}\Bigr|}-\ii\pi\,\mathrm{sgn}(E)\Theta(D-|E|)\Bigr]\,,
\end{equation}
with the graphene bandwidth $D=\sqrt{\sqrt{3}\pi}t\simeq 6$~eV, see for example Refs.~\onlinecite{Wehling2010:PRL,Kochan2014:PRL}.

\begin{figure}
\includegraphics[width=0.98\columnwidth]{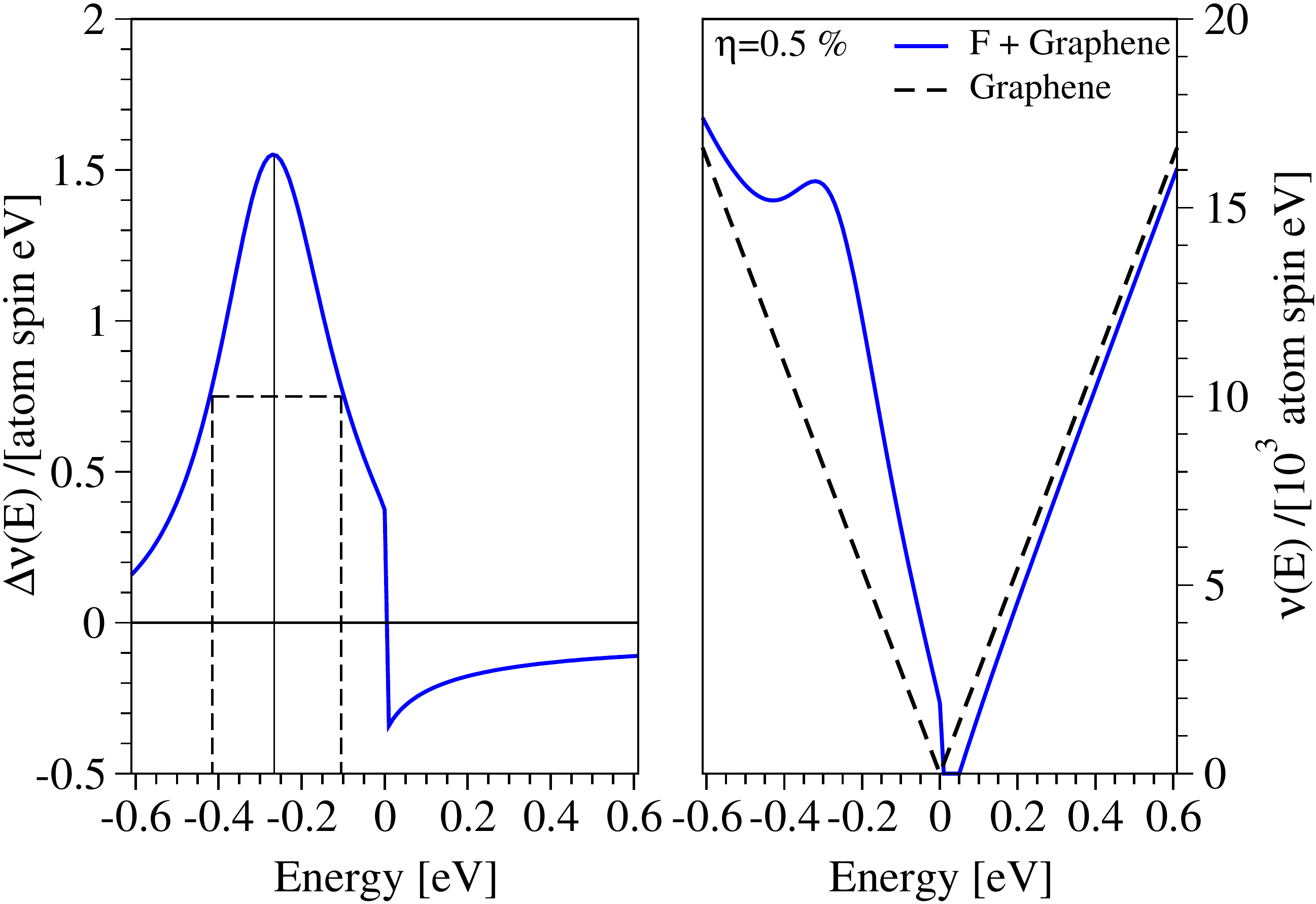}
\caption{\label{Fig:DOS-model-calculation}(Color online) Left panel: change $\Delta\nu(E)$ in DOS per atom and spin, see Eq.~(\ref{Eq:DOS-change}), computed for the single-impurity model employing orbital tight-binding parameters $T=5.5$~eV and $\varepsilon_{\mathrm{f}}=-2.2$~eV. The pronounced maximum appears at $E\simeq -0.26$~eV with FWHM$\ \simeq 0.3$~eV. Right panel: perturbed DOS per atom and spin $\nu(E)=\nu_0(E)+\eta\,\Delta\nu(E)$ as functions of energy for the fluorine impurity concentration $\eta=0.5\%$ (solid line). The dashed line shows the unperturbed DOS per atom and spin near the graphene neutrality point.}
\end{figure}

Our analytical results are shown in Fig.~\ref{Fig:DOS-model-calculation}, in which we plot
$\Delta\nu(E)$ as well as the full DOS per atom and spin, $\nu(E)=\nu_0(E)+\eta\,\Delta\nu(E)$,
as functions of energy, for fluorinated graphene at very low concentrations. The quantity $\nu_0(E)$ describes the DOS per atom and spin of the unperturbed graphene, $\nu_0(E)=-\frac{1}{\pi}\,\mathrm{Im}\,G_0(E)$. To visualize the changes
in DOS we take an unrealistically large concentration (whose effect is purely multiplicative) of $\eta = 0.5\%$.
From Fig.~\ref{Fig:DOS-model-calculation} we see that fluorine dominantly affects the DOS at energies around
$E\simeq -0.26$~eV which corresponds to a pronounced peak in $\Delta\nu(E)$ with the FWHM$\ \simeq 0.3$~eV.
This gives the energy window $(-0.4\,\mathrm{eV},-0.1\,\mathrm{eV})$ where we expect dominant orbital effects of fluorine chemisorption; see also results of Ref.~[\onlinecite{Ihnatsenka2011:PRB}]. There is a recent experimental evidence,\cite{Tahara2013:APL} based on electron-hole asymmetry in transport, indicating resonant scattering due to fluorination, but further studies are certainly
called for.

\section{Conclusions}

We studied, from first principles, the electronic structure and spin-orbit splitting of spin unpolarized fluorinated
graphene. Two important limits were covered: a dense limit, represented by a $1 \times 1$ supercell with a single
fluorine in the top position in the cell, and a dilute limit, represented by a $10 \times 10$ supercell, also with
a single top-positioned fluorine adatom. We further looked at the intermediate case of a $5 \times 5$ supercell,
to contrast differences with the case of hydrogenated graphene. All our investigated structures were structurally
relaxed.

For the dilute limit we introduced a single-orbital tight-binding (hopping) Hamiltonian that very nicely reproduces
the first-principles results. The orbital part of the Hamiltonian is based on the changes of the adatom energy
and includes also hopping between the fluorine and fluorinated carbon. The spin-orbit part includes the usual
intrinsic (spin-conserving next-nearest neighbor) and Rashba (spin-flip nearest neighbor) hoppings, as
well as new PIA hoppings which describe spin-flip paths between next-nearest neighbors. We give a specific
description of the Hamiltonian and the complex phases of the hoppings, as well as best fits to the DFT obtained
spin-orbit spin splittings. We also use our tight-binding Hamiltonian to investigate superlarge supercells ($40 \times
40$) to search for band structure resonances due to fluorine adatoms. Using nonperturbative analytical
calculations we also obtain what would be a single-adatom representation of the changes in the density of states.
We believe that our tight-binding model Hamiltonian is reliable to explain the physics near the Fermi level of fluorinated graphene and can be used in quantum transport simulations that involve orbital and spin-orbit effects, say
momentum and spin relaxation, charge and spin transport, or the spin Hall effect.

We draw several conclusions from our investigations: (i) Fluorine induces a giant spin-orbit coupling, an order
of magnitude greater than hydrogen. This is evidenced directly by the dense limit results, but also by the
magnitudes of the obtained fits to our tight-binding model. (ii) The enhancement of spin-orbit coupling is
not principally due to the $\sigma-\pi$ hybridization induced by structural deformation, as in the case
of hydrogen. In fact, the giant enhancement of SOC on graphene band structure due to fluorine adatoms
comes from the native spin-orbit coupling of fluorine, and its orbital hybridization with carbon. (iii) Fluorine
adatoms are only weak or marginal resonant scatterers. The resonant peak in the density of states lies
260 meV below the Dirac point. The peak is about 300 meV broad. This again contrasts with hydrogen
adatoms which are perfect examples of narrow resonance scatterers at the Dirac point.

\begin{acknowledgments}
We thank Jun Zhu for stimulating discussions. This work was supported by the DFG SFB 689 and
GRK 1570, and by the European Union Seventh Framework
Programme under Grant Agreement No. 604391 Graphene Flagship.
\end{acknowledgments}

\appendix
\section{Intermediate limit: $5 \times 5$ supercell}
\label{sec:5-times-5}
\begin{figure}[htp]
\includegraphics[width=0.98\columnwidth]{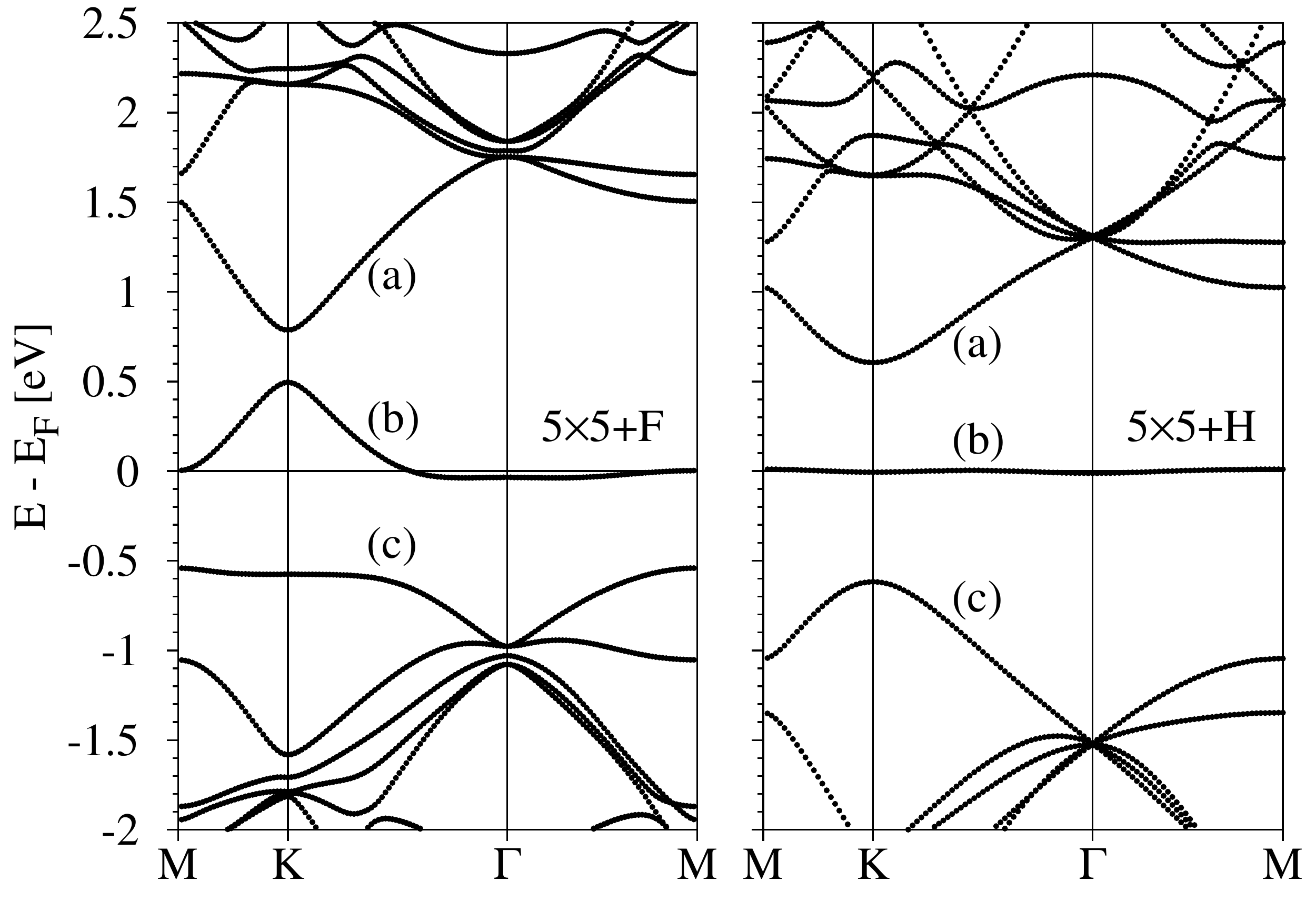}
\caption{\label{fig:GrFH5}Calculated electronic band structures of fluorinated (left panel) and hydrogenated (right panel) graphene
within $5\times 5$ supercell configurations. Three bands (a), (b), and (c) correspond to the conduction, midgap, and valence band, respectively, as in the main text. Formation of the midgap state around the Fermi level dominates both band structures. However, for hydrogenated graphene this band is hardly dispersive, giving the pronounced resonance at
the Dirac point.}
\end{figure}

In what follows we summarize our findings for the intermediate limit of a $5\times 5$ supercell, which lies in
between the dense and dilute ones. For hydrogenated graphene \cite{Gmitra2013:PRL} such
a supercell is already large enough to represent dilute coverage. Our $5\times 5$ supercell is fully relaxed,
with the lattice constant $12.297$~\AA, which differs by $0.2\permil$ from the same-size pure graphene supercell lattice constant ($12.3$~\AA). Bader charge analysis shows that fluorine acquires a
negative charge of 0.518~e. In contrast, for the hydrogenated case there is almost zero charge
transfer to hydrogen. In Fig.~\ref{fig:GrFH5} we show the calculated electronic band structures for the fluorinated and hydrogenated graphene for comparison. In both cases one sees a formation of the midgap states around the Fermi level.
However, the band width of 0.6~eV for the fluorinated graphene is much larger than for hydrogenated graphene,
for which the bandwidth is less than 0.1~meV.\cite{Gmitra2013:PRL}

\begin{figure}[htp]
\includegraphics[width=0.98\columnwidth]{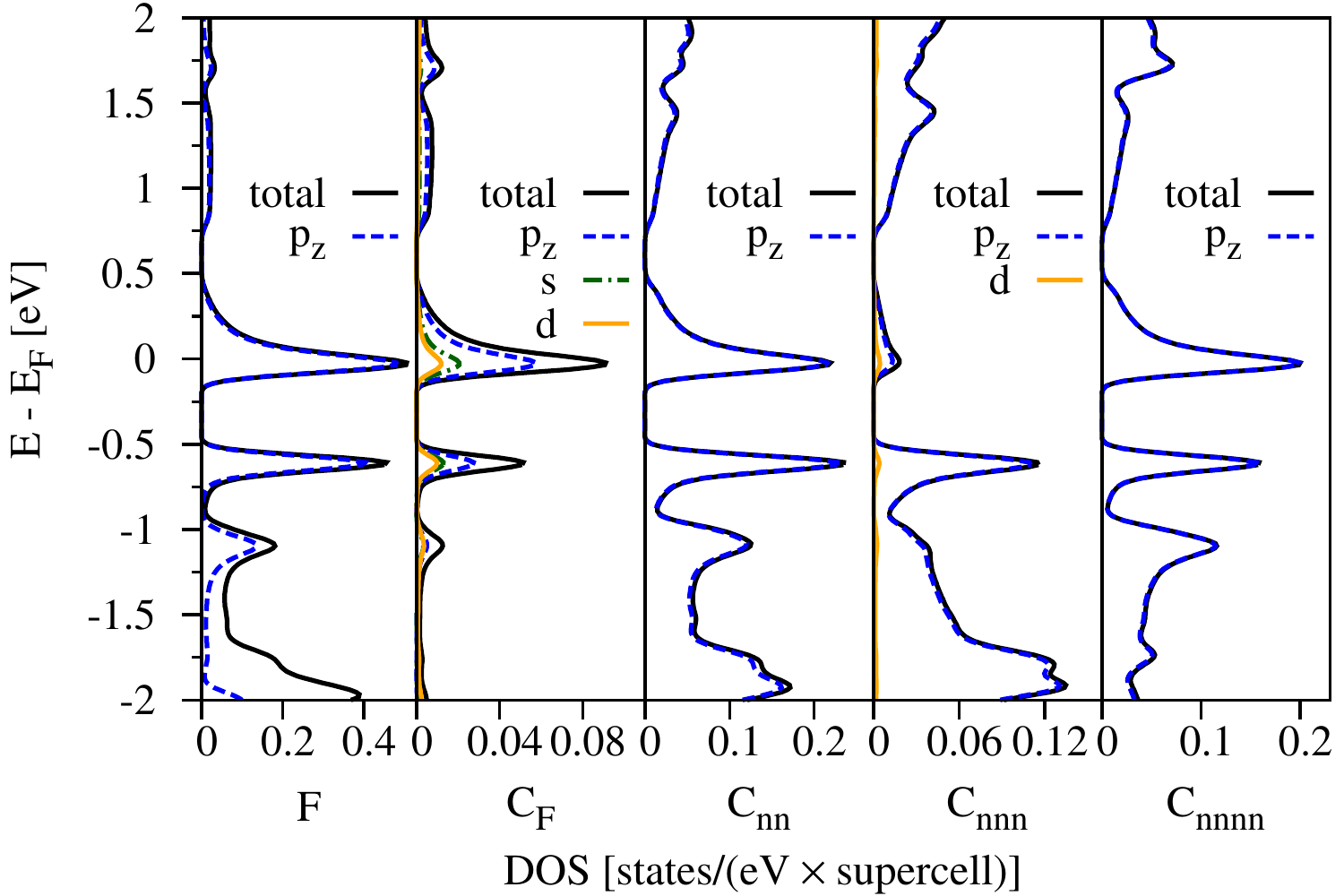}
\caption{\label{fig:GrF5_pdos}(Color online) Calculated orbital resolved atomic DOS for fluorinated graphene in a $5\times 5$ supercell.
The orbital resolved DOS is shown for the fluorine atom $\mathrm{F}$, fluorinated carbon C$_\text{F}$, and its
nearest, next-nearest, and next-next-nearest neighbors, C$_\text{nn}$, C$_\text{nnn}$, and C$_\text{nnnn}$, respectively.}
\end{figure}
\begin{figure}[htp]
\vspace{1mm}
\includegraphics[width=0.98\columnwidth]{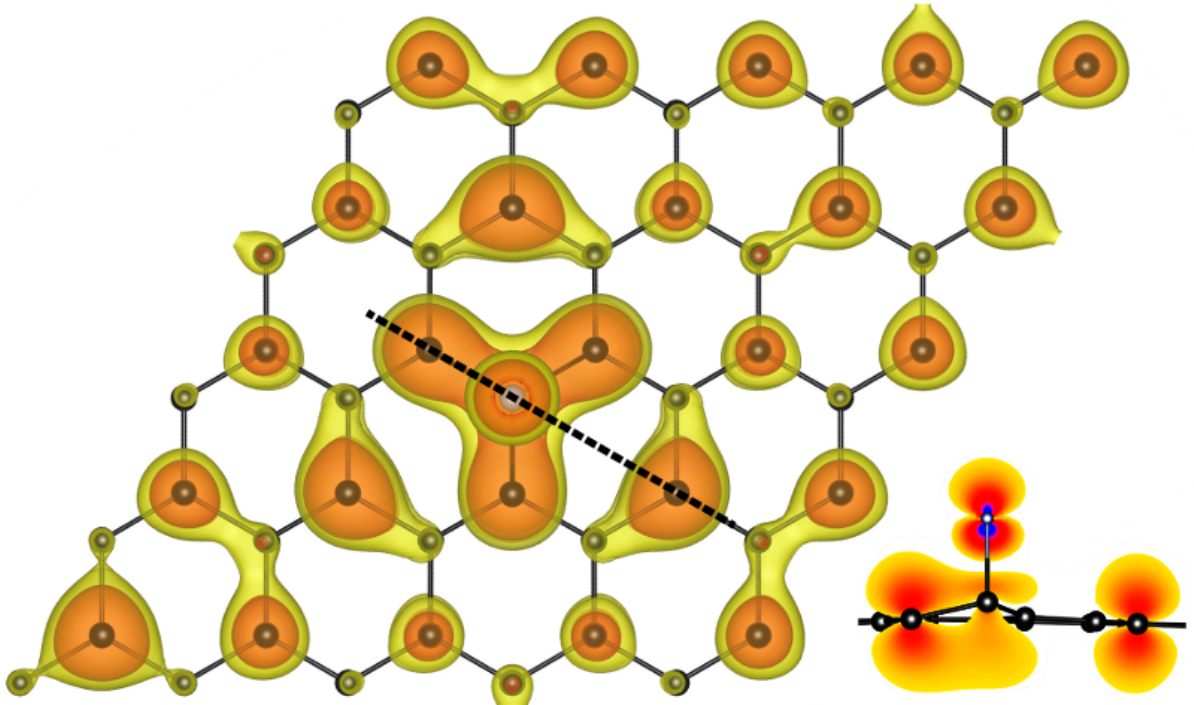}
\caption{\label{fig:GrF5_density}(Color online) Top view of the valence charge density plot of fluorinated graphene on a
$5\times5$ supercell.
The charge density was obtained by summing the absolute squares of the Kohn-Sham states lying in the energy interval between $-0.2$~eV and $0.6$~eV.
The dark (orange) surface is the isovalue of 0.002~$\text{\AA}^{-3}$ and the light (yellow) one corresponds to 0.001~$\text{\AA}^{-3}$.
Dashed lines represent the cross-sectional view shown in the bottom right.}
\end{figure}

The DOS for selected atoms in the vicinity of fluorine are shown in Fig.~\ref{fig:GrF5_pdos}. Orbital resolved analysis shows a significant contribution of $p_z$ orbitals to the states in the studied
energy window around the Fermi level. The spatial distribution of the $p_z$ orbitals can be visualized by the electronic density $\rho(\mathbf{r}) = \sum_{n,\textbf{k}} |\phi_{n}^\textbf{k}(\mathbf{r})|^2$.
In Fig.~\ref{fig:GrF5_density} we show a top view of the electronic density summed over the Kohn-Sham eigenstates
$\phi_{n}^\textbf{k}$ including eigenstates with energies $\varepsilon_{n}^\mathbf{k}$ lying within
the energy window $\varepsilon_\text{min}=-0.2$~eV and $\varepsilon_\text{max}=0.6$~eV with respect to the Fermi level. The dashed line in the main figure corresponds to the cross-sectional view shown in Fig.~\ref{fig:GrF5_density}.
From the plot one sees that the midgap state represents a strongly delocalized state mostly situated on the ``nonfluorinated'' sublattice (i.e.,~sublattice B, fluorinated carbon $\mathrm{C_F}$ is on sublattice A).
The appearance of such a delocalized state indicates that a $5\times 5$ supercell is not large enough to represent the dilute fluorination limit, since there is still significant fluorine interaction among the periodic images.

We find that sublattice B is the main contributor to the midgap DOS (80\%). The fluorine atom and the carbon atoms at sublattice A account for only 10\% each.
In the valence band at $-0.6$~eV, sublattice B represents 54\%, sublattice A 36\%, and fluorine 10\% of the local DOS.
This decomposition can be seen in Fig.~\ref{fig:GrF5_pdos} where the orbital resolved DOS for $\mathrm{F}$, C$_\text{F}$, and all carbon atoms (C$_\text{nn}$, C$_\text{nnn}$, and C$_\text{nnnn}$) up to the third nearest neighbors of C$_\text{F}$ are provided.

A large---predominantly $p_z$---contribution to the midgap states comes from the fluorine atom with 0.5~states/eV;
see the first panel in Fig.~\ref{fig:GrF5_pdos}. Atoms C$_\text{nn}$ and C$_\text{nnnn}$ belonging to the sublattice B, both carry mainly $p_{z}$ character and contribute equally
to the valence band edge at $-0.6$~eV and the midgap level. Their DOS is very similar in shape and value.
The DOS at fluorinated carbon C$_\text{F}$ has a more complex composition (visible also in the inset of
Fig.~\ref{fig:GrF5_density}), where $p_z$, $s$, and $d$ characters are identified, but they sum up only to 0.09~states/eV. The overall contribution of C$_\text{F}$ within the energy region from $-2$~eV to $2$~eV is very low.
The DOS contribution from C$_\text{nnn}$ carbons at sublattice A is negligible for the midgap and the valence band edge and also possesses mainly $p_z$ character. The conduction band at energies above $0.8$~eV has mainly $p_z$ character for all the atoms analyzed in Fig.~\ref{fig:GrF5_pdos}.

\begin{figure}[htp]
\includegraphics[width=0.98\columnwidth]{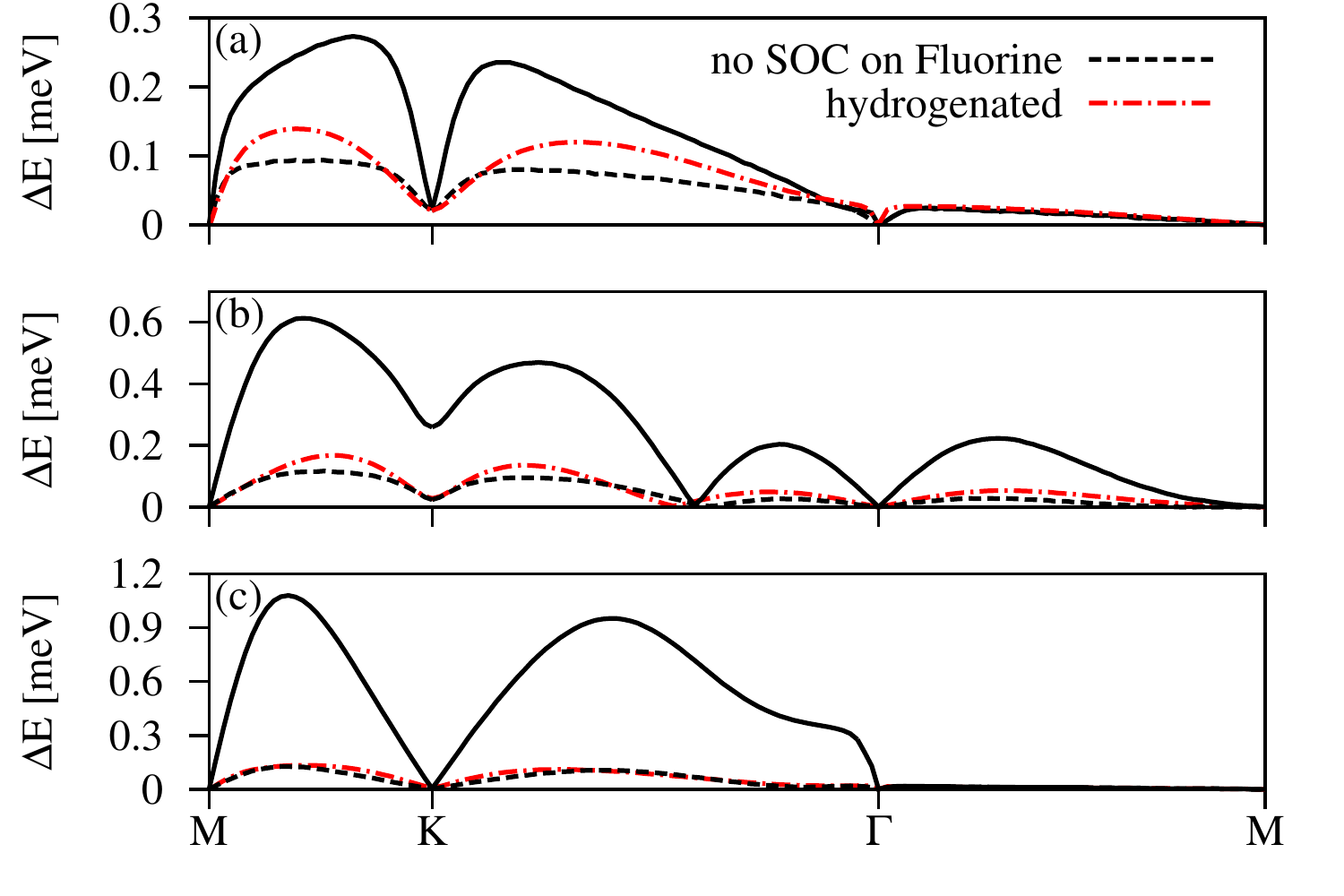}
\caption{\label{fig:GrF5_soc_splitting}(Color online) SOC splittings for the conduction (a), midgap (b), and valence (c) bands with respect to the Fermi level
 of the intermediately fluorinated $5\times 5$ system. Dashed lines are representing the SOC splittings without the intra-atomic SOC contributions from
 the F atom. For comparison the SOC splittings of the hydrogenated $5\times 5$ system are provided by the dashed-dotted lines.}
\end{figure}

Finally, Fig.~\ref{fig:GrF5_soc_splitting} shows the absolute values of spin-orbit splittings for the conduction (a),
midgap (b), and valence (c) bands. The largest splittings range from 0.26~meV for the conduction band, via 0.6~meV
for the midgap band, up to 1.1~meV for the valence band. If we turn off the intra-atomic SOC on fluorine, the splittings
are drastically reduced, essentially to the hydrogenated graphene level. This nicely demonstrates the limit of
what spin-orbit splittings can be achieved by $sp^3$ bonding. If a larger splitting is observed, the spin-orbit coupling comes most likely from the adatom itself, not from the host graphene lattice.


\bibliography{flgr}


\end{document}